\documentclass[%
 reprint,
 superscriptaddress,
 showpacs,preprintnumbers,
 showkeys,
 nofootinbib,
 amsmath,amssymb,
 aps,
]{revtex4-1}

\usepackage{graphicx}%
\usepackage{physics}
\usepackage{subfiles}
\usepackage[caption=false]{subfig}

\begin{document}

\title{Restricted Boltzmann Machines for galaxy morphology classification with a quantum annealer}%

\author{Jo\~ao~Caldeira}
\thanks{J.~Caldeira and J.~Job contributed equally to this work}
\email{caldeira@fnal.gov}
\email{joshua.job@lmco.com}
\affiliation{Fermi National Accelerator Laboratory, Batavia, IL 60510}

\author{Joshua~Job}
\thanks{J.~Caldeira and J.~Job contributed equally to this work}
\email{caldeira@fnal.gov}
\email{joshua.job@lmco.com}
\affiliation{Lockheed Martin Advanced Technology Center, Sunnyvale, CA 94089}

\author{Steven~H.~Adachi}
\affiliation{Lockheed Martin Advanced Technology Center, Sunnyvale, CA 94089}

\author{Brian~Nord}
\affiliation{Fermi National Accelerator Laboratory, Batavia, IL 60510}
\affiliation{Kavli Institute for Cosmological Physics, University of Chicago, Chicago, IL 60637}
\affiliation{Department of Astronomy and Astrophysics, University of Chicago, Chicago, IL 60637}

\author{Gabriel~N.~Perdue}
\affiliation{Fermi National Accelerator Laboratory, Batavia, IL 60510}

\date{\today}%

\begin{abstract}
We present the application of Restricted Boltzmann Machines (RBMs) to the task of astronomical image classification using a quantum annealer built by D-Wave Systems.
Morphological analysis of galaxies provides critical information for studying their formation and evolution across cosmic time scales. 
We compress galaxy images using principal component analysis to fit a representation on the quantum hardware.
Then, we train RBMs with discriminative and generative algorithms, including contrastive divergence and hybrid generative-discriminative approaches, to classify different galaxy morphologies. 
The methods we compare include Quantum Annealing (QA), Markov Chain Monte Carlo (MCMC) Gibbs Sampling, and Simulated Annealing (SA) as well as machine learning algorithms like gradient boosted decision trees. 
We find that RBMs implemented on D-Wave hardware perform well, and that they show some classification performance advantages on small datasets, but they don't offer a broadly strategic advantage for this task. 
During this exploration, we analyzed the steps required for Boltzmann sampling with the D-Wave 2000Q, including a study of temperature estimation, and examined the impact of qubit noise by comparing and contrasting the original D-Wave 2000Q to the lower-noise version recently made available.
While these analyses ultimately had minimal impact on the performance of the RBMs, we include them for reference.

\end{abstract}

\preprint{FERMILAB-PUB-19-546-QIS-SCD}

\pacs{Valid PACS appear here}%

\maketitle

\section{Introduction}
\label{sec:introduction}

Machine learning techniques are being used increasingly in high energy physics \cite[e.g.,][]{Albertsson2018} and astrophysics \cite[e.g.,][]{Ntampaka2019} for applications such as event detection, particle identification, data analysis, and the simulation of detector responses.  
In many cases, machine learning provides an efficient alternative to analytical models, which are often intractable, or Monte Carlo-based simulations, which can be computationally expensive.
Data analysis tasks in cosmology are extremely computing intensive and will become even more so as new instruments like LSST \cite{lsst} come online, motivating advances in data analysis techniques.

Feynman  first proposed the idea of using quantum computers to simulate physical systems \cite{Feynman1982}.
More recently, a variety of approaches have been studied for combining quantum computing with machine learning techniques \cite{Biamonte2017}.
While quantum computing hardware is still in the early stages of development, initial attempts have been made to apply quantum machine learning to high energy physics, e.g.\ classifying Higgs decay events in Large Hadron Collider data \cite{Mott2017}.

In this study, we focus on the challenge of morphological classification of galaxies via astronomical images using the D-Wave 2000Q quantum annealer \cite{dwave2000q}. 
Galaxies exhibit morphologies (structure in their shape) that tend to correlate with their evolutionary state and history. 
For example, spiral galaxies (typically blue) have higher rates of star formation often visible in clumpy regions of their spiral arms, irregulars have quasi-randomly distributed clumps of star formation, and elliptical galaxies (typically red) tend to have ceased making their stars.
Star formation occurs more readily in relaxed kinematic environments, where gravity has sufficient relative influence to pull together cold material that can fuse into stellar cores. 
Highly energetic or dense environments, like the cores of galaxy clusters or filaments of the cosmic web, may cause galaxy mergers or other disruptive events that can slow or halt star formation. 
Galaxies evolve rapidly in stellar mass in the range of cosmic redshifts (measures of cosmic age) $1 < z < 3$, where star formation density peaks near $z\sim 2.5$.
The rate of star formation is one of the primary measures of cosmic energy exchange, and structural and morphological analysis of galaxies permits a critical avenue of investigation of cosmic evolution. 
The accurate classification of galaxies based on morphology is a critical step in this analysis. 
Please see the review in Ref.~\cite{Conselice2003} for more details. 

Classical methods for morphological analysis have typically relied on a) visual examination, such as conducted through the Galaxy Zoo project \cite{Willett2013}; multi-wavelength model-fitting \cite{vika2015}; and structural proxies, like concentration, asymmetry, and clumpiness \cite{Conselice2003}. 
Recent advancements in deep learning permit the usage of convolutional neural networks for morphological classification, which have become the state of the art \cite{Dieleman2015,Tuccillo2017,Barchi2019}.
The conventional convolutional neural network does not yet have an efficient implementation on the D-Wave quantum annealer. 
In this work, we use a different type of machine learning model, the restricted Boltzmann machine (RBM) \cite{Smolensky1986}.
While there are many other types of machine learning models, the RBM model has stochastic binary variables and a quadratic energy functional, which can be efficiently implemented using the relatively small number of qubits available on near-term quantum computing devices, such as the D-Wave quantum annealer \cite{Johnson2011,Pudenz2013}.
Training an RBM is classically hard, but there is reason to believe quantum annealers may eventually offer performance advantages \cite{Adachi2015}.

While quantum annealers are generally used for solving optimization problems, they have also been used in a machine learning context, where the quantum annealer is programmed with coefficients derived from the RBM, and used as a sampling engine to generate samples from the Boltzmann distribution \cite{Adachi2015,Benedetti2016}.
As in classical machine learning, an iterative training process is used to refine the RBM coefficients.

Quantum annealers offer a way to leverage the power of quantum computers while avoiding the complexity of a gate-based programming model, making them an attractive tool for domain scientists.
However, given the limitations of present-day quantum annealers, this approach presents a number of challenges.
For instance, input data must be severely compressed to fit the available qubits.
Further, samples from the quantum annealer may be not be Boltzmann distributed, in which case post-processing or temperature estimation techniques may need to be applied.

\subsection{Overview of the paper}
\label{sec:paperplan}

This paper consists of two main results.
First we show the outcome of some studies of the distribution of states coming from the D-Wave. 
We considered a variety of post-processing techniques to bring the output distributions closer to Boltzmann distributions, which are theoretically necessary for training RBMs.
Second, we show the results of trained RBMs and other algorithms for the galaxy morphology classification problem.
We also include a discussion of the data and compression methods employed.

Specifically, in Section \ref{sec:datacompression} we briefly review the galaxy morphology classification datasets used for training and testing, and the techniques used to compress the data for the D-Wave 2000Q.
In Section \ref{sec:generativediscriminative} we discuss the training algorithms used to prepare RBMs for classification.
We compare a variety of options for training RBMs, including various combinations of generative and discriminative training.
In Section \ref{sec:howtoboltzmann}, we discuss a variety of post-processing steps for producing a Boltzmann-distributed set of energy states using the D-Wave quantum annealer.
We also compare two versions of the D-Wave 2000Q, one of which featured lower levels of noise.
In Section \ref{sec:results} we study the performance of RBMs on the quantum device and using classical resources, and we compare the performance of other classical machine learning algorithms.
Finally, in Section \ref{sec:conclusions} we conclude and offer thoughts on future directions.

\section{Astronomical Data and Compression}
\label{sec:datacompression}

\subsection{Data: Galaxy Zoo}

We use data from the Galaxy Zoo 2 data release, which contains 304,122 galaxies taken from the Sloan Digital Sky Survey \cite{Willett2013}.
We use the subset of this data in the training set for the Galaxy Zoo Kaggle challenge,\footnote{\url{https://www.kaggle.com/c/galaxy-zoo-the-galaxy-challenge}} which contains 61,578 galaxies.
For each image of a galaxy, this dataset includes crowdsourced answers to a set of 11 questions characterizing the galaxy's morphology.
These include the presence or absence of different features like bulges, disks, bars, and spirals. 
We simplify the problem into a binary classification problem by picking spiral galaxies (those with more than 50\% ``yes'' answers to ``Is there a spiral pattern?'') and rounded smooth galaxies (those with more than 50\% ``completely round'' answers to the question ``How rounded is it?'').
These classes contain 10,397 and 8,434 galaxies, respectively.
We select 5,000 random images of each of the two classes.
Before applying any data compression algorithm, we crop the images to 200 by 200 pixels.

\subsection{Compression and Manipulation}

Raw images are $200\times 200$ RGB pixels in size, which is far too large to encode in the binary variables available on the D-Wave 2000Q.
There are a number of interesting compression schemes available, including, for example, discrete variational autoencoders \cite{2016arXiv160902200R,2018arXiv180507445V}.
In practice, we found no appreciable advantage between different compression schemes while reducing data dimensionality to the level where we could encode the essential information about a given image into the binary variables available.
Therefore, we relied on principal component analysis (PCA) on the basis that the method is simple and easy to explain and understand.

\begin{figure}[htp]
\centering
\includegraphics[width=0.99\linewidth]{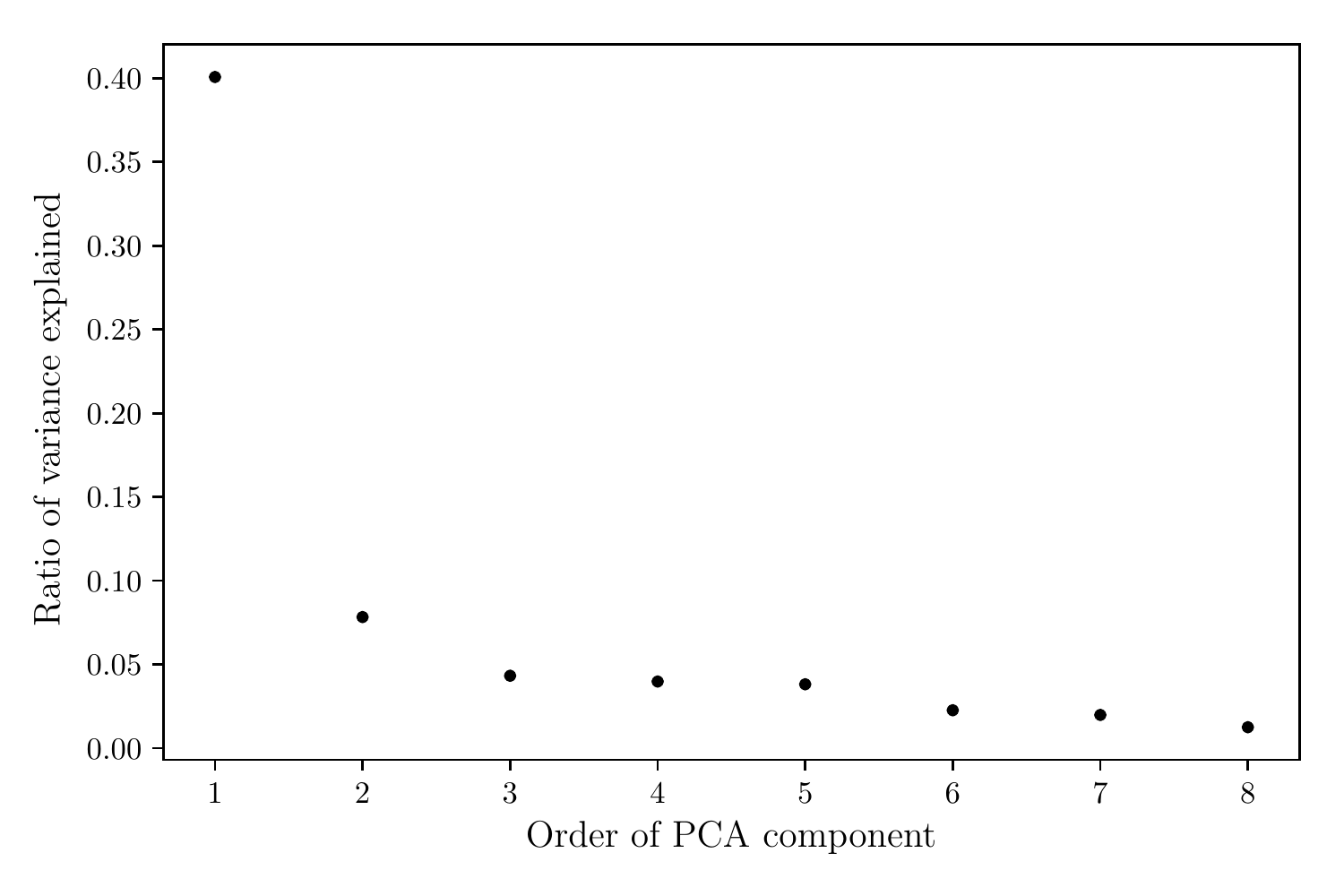}
\caption{The ratio of the total variance in the PCA training dataset explained by each additional PCA component.}
\label{fig:pca_variance_explained}
\end{figure}

We used 5,000 images to train a PCA model using Scikit-learn \cite{scikit-learn}, and applied it on the remaining 5,000 images to obtain the dataset we used to train and test the RBM.
The ratio of explained variance added by each PCA component is shown in Fig.~\ref{fig:pca_variance_explained}. We can see that the information contained in each additional component rapidly decays.

\begin{figure}[htp]
\subfloat[Galaxies with label 0, corresponding to rounded smooth galaxies.]{\label{fig:galaxymorphology_batch004_catrounded}%
  \includegraphics[width=0.45\textwidth]{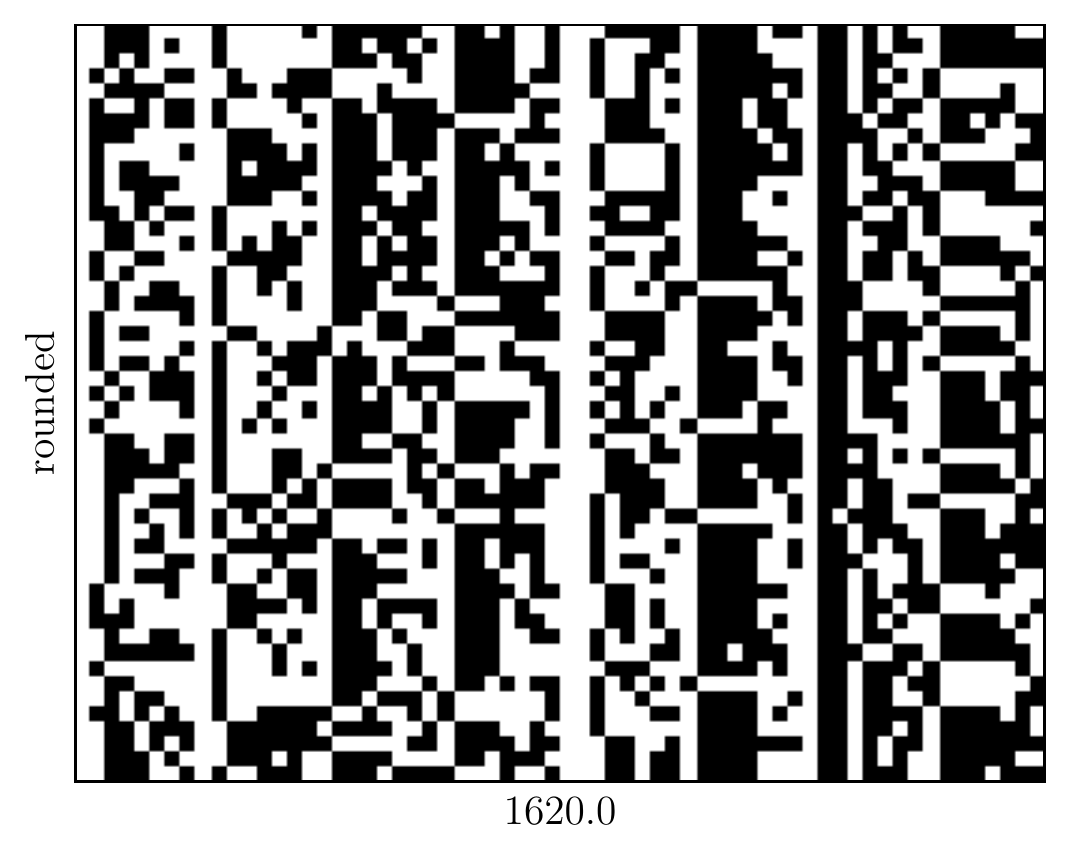}%
} \\
\subfloat[Galaxies with label 1, corresponding to spiral galaxies.]{\label{fig:galaxymorphology_batch004_catspiral}%
  \includegraphics[width=0.45\textwidth]{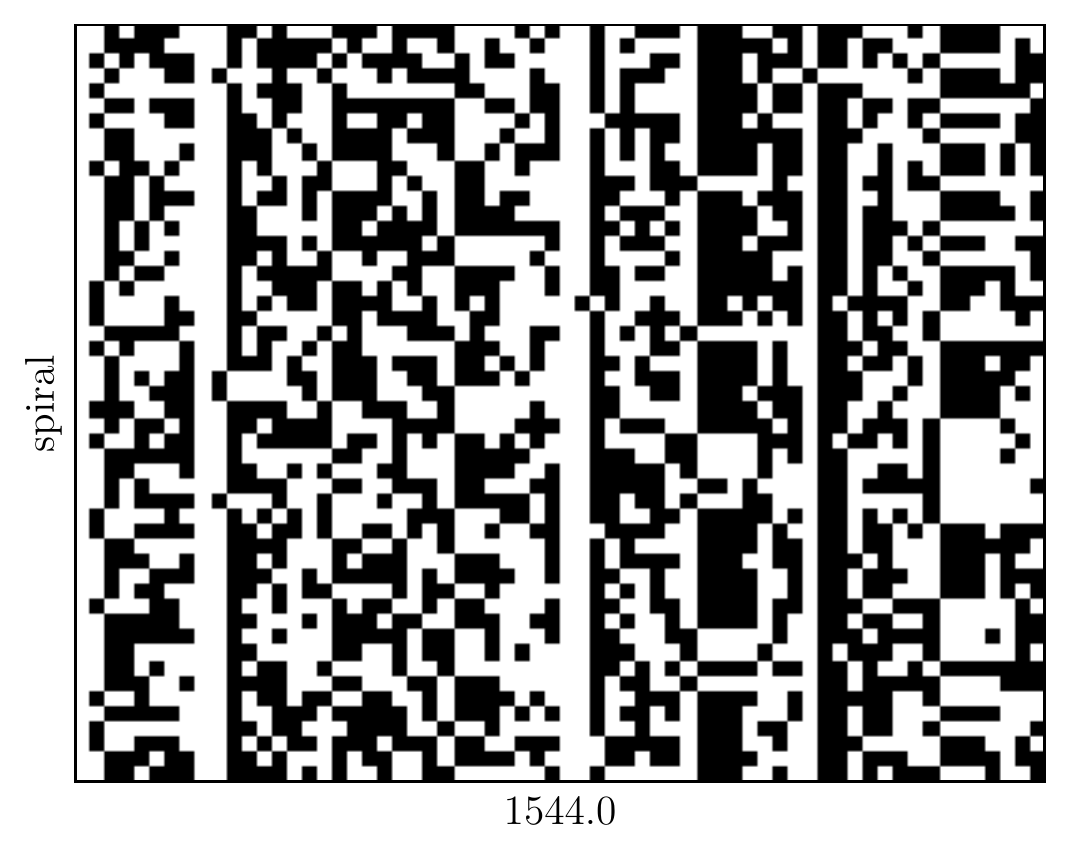}%
}
	\caption{
	Compressed ``minibatches'' of data. Here each row in each figure represents one galaxy image. 
	There are fifty events (rows) in each figure.
	Each column represents the binary value of the compressed data bit, with dark colors indicating zero and bright colors indicating one.
	Here we have 64 bits of width, with each PCA component represented by an 8-bit discretization of the floating point value, for a total of eight PCA components represented, with the leading component on the left side of the figure.
	The x-axis labels report the sum of all the binary values across the entire minibatch (so the maximum possible is 3,200).
	}
	\label{fig:compressed_morphology_minibatches}
\end{figure}

The encoded components defined by PCA are 64-bit floating point numbers.
We would like to transform them into a more compact representation.
We will do this by linearly mapping the range of each PCA component in the training set to the interval [15, 240].
We can then round to the nearest integer and transform the data into unsigned 8-bit integers, which support numbers between 0 and 255.
The range we map the training set into was chosen to be safely inside the [0, 255] interval in order to accomodate outliers beyond the ranges present in the training set.
See Fig. \ref{fig:compressed_morphology_minibatches} for a visualization of a sample of compressed data.

To these compact representations of the images, we add a bit representing the class (0 for rounded smooth galaxies and 1 for spiral galaxies).
This means that if our RBM has $n$ visible units, the first $n-1$ will correspond to the first bits in the compressed images, while the last visible unit will encode the class of each image.

During the analysis we were concerned that the digitization scheme employed was putting extra weight on the most significant bits of each encoded PCA component, but this was not information we could easily share directly with the RBM algorithm.
We tested several different re-ordering schemes for the bits and also tested preferentially keeping the most significant bits only for higher order components as a way of including information from those components when working with small RBMs.
However, the re-ordering schemes generally slightly degraded performance, and attempts to include a larger number of components using only the most significant bits did not offer any performance advantages.

\section{Generative and Discriminative Training}
\label{sec:generativediscriminative}

\subsection{RBM training}
Restricted Boltzmann machines (RBMs) are generally speaking generative models, where one attempts to approximate a target distribution over a string of binary variables $\vec{v}$, $p(\vec{v})$, as the marginal distribution of a larger system $p(\vec{v})=\sum_{\vec{h}} p(\vec{v},\vec{h})$ composed of binary variables $\vec{v}$ and latent variables $\vec{h}$, with an ansatz such that 
\begin{equation}
p(\vec{v},\vec{h}) \propto \exp(-v'W h + b'v + c'h) \label{eqn:rbmprobdistribution}
\end{equation}
for some bias vectors $\vec{b}$, $\vec{c}$, and a weight matrix $W$.
This corresponds to a complete bipartite graph with local biases on the nodes and interactions along the edges.
The binary variables $\vec{v}$ are called the visible nodes, as they compose the distribution of interest, while the $\vec{h}$ are the hidden nodes.

By variationally maximizing the log-likelihood of the data of the RBM model with respect to the weights and biases, we can train the RBM to better approximate the data distribution.
It can be readily shown that maximizing the log-likelihood corresponds to matching the one and two-point correlation functions of the model between states conditioned on the data distribution and the free generative model.
Defining the loss as the negative log-likelihood, dubbed $L$, we get derivatives for the variational parameters
\begin{align}
\pdv{L}{b^i}&=\expval{v^i}_{\text{data}} - \expval{v^i}_{\text{model}} \\
\pdv{L}{c^i}&=\expval{h^i}_{\text{data}} - \expval{h^i}_{\text{model}} \\
\pdv{L}{W_i^j}&=\expval{v^i h^j}_{\text{data}} - \expval{v^i h^j}_{\text{model}}
\end{align}
Collectively these derivatives form the gradient, to be used in gradient descent to the adjustments for $\vec{b}$, $\vec{c}$, and $W$ respectively.
Here the expectations are computed over the training set and the the model (also called the positive and negative phases).

Once the RBM has been trained, we can use it to make a prediction on the class of unseen images.
To do this, we calculate the free energies of the RBM with the visible units set to the compressed image representation and both options for the class.
The class corresponding to the lowest free energy is then the most likely class for that image.
This type of discriminative RBM was introduced in Ref.~\cite{Larochelle2008}.

\subsection{Classical training algorithms}

In general, one cannot compute expectations over the model directly, as it takes a time that scales as $2^{\min{n_v,n_h}}$ where $n_v$ and $n_h$ represent the number of visible and hidden units, respectively.
This is generally intractable.
However, we can use a variety of algorithms to perform training.

\subsubsection{Contrastive divergence}

We may perform efficient block sampling updates of $p(v|h)$ and $p(h|v)$, as the conditional distributions reduce to single-spin probabilities that may be sampled in linear time with the number of variables.
Initializing a Markov chain performing such block Gibbs sampling at each training datapoint and taking expectations over the resulting chains is the basis of the contrastive divergence (CD) algorithm, first put forward in Ref.~\cite{Hinton2002}.
Using CD, one can often train RBMs of quite large size reasonably efficiently.

\subsubsection{Discriminative training}

In this work, we are interested in using RBMs not as a strictly generative model, but as a classification algorithm.
In essence, we wish to be able to input an image and sample the posterior distribution for the class of that image using the RBM.
Thus, rather than directly modeling the full $p(\vec{v})$, as is standard practice, we are really interested in only $p(v_{\text{class}}|\vec{v}_{\text{image}})$.
Rather than training a model to represent the entire distribution over $\vec{v}$, we can instead directly train to maximize the log-likelihood of the $p(v_{\text{class}}|\vec{v}_{\text{image}})$ distribution, as was proposed in Ref.~\cite{Larochelle2008}.

The training process is much the same as before, except that now the sample over the model is vastly simplified, as one is taking an expectation with the image dataset fixed, reducing the effective number of variables to merely that of the number of bits used to represent the class.
In our case, where we use a single variable, we thus can contract the graph in linear time to get an exact gradient.
In general, one can contract in a time scaling no worse than $2^{n_{\text{class}}}$ for unary encoding of the classes.
Using a binary representation for the classes, one can do this in linear time in the number of classes, and thus training the discriminative model can be done efficiently on a classical computer in an exact fashion, with no Markov chains required.

\subsubsection{Hybrid approaches}

Finally, one may consider a hybrid approach.
For instance, an approach where one takes a combination of both the aforementined gradients, generative and discriminative, so as to better approximate $p(v_{\text{class}}|\vec{v}_{\text{image}})$ while still representing the full distribution efficiently.
In this, we can set a value $\lambda$ which combines the the gradients $\grad_{\text{gen}}$ and $\grad_{\text{disc}}$ for the generative and discriminative models as 
\begin{equation}
\grad_{\text{hybrid}} = \frac{\lambda}{1+\lambda}\grad_{\text{gen}}+\frac{1}{1+\lambda} \grad_{\text{disc}}.
\end{equation}
This approach was also investigated in Ref.~\cite{Larochelle2008} and found to be beneficial at small values of $\lambda$.

We additionally investigate another hybrid approach, where we use generative training as a kind of pretraining and then follow it with pure discriminative training, which we dub ``annealed hybrid'' training, even though if one is considering it as annealing the $\lambda$ parameter it is better thought of as a quench.
This was motivated by our observations of the performance of generative and discriminative training.

\subsection{Generative training with quantum annealing}

Finally, we compare classical training algorithms against a quantum annealing (QA) based model for estimating the negative phase (the intractable model expectation values) and alternatives to QA, including a pure Gibbs sampling MCMC algorithm initialized at a random position, and simulated annealing.
In essence we seek to understand what causes observed QA performance by testing against other annealing algorithms.

\begin{figure*}[htp]
\includegraphics[width=0.5\textwidth]{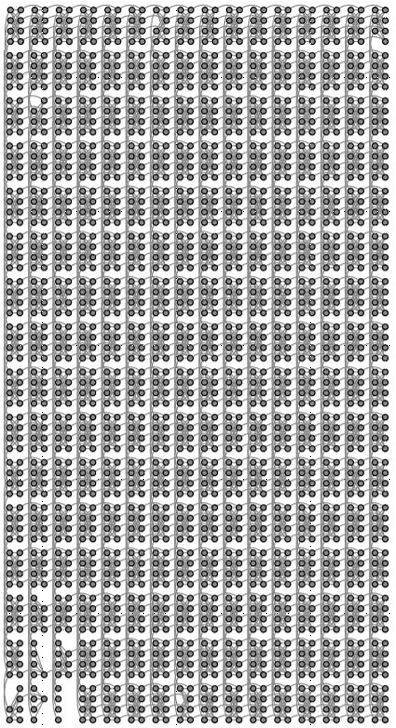}
\centering
\caption{An example Chimera lattice, from the low-noise DW2000Q we used.
Each line is a coupler, each node an active qubit. The ideal graph is a tile of $K_{4,4}$ graphs with vertical connections between qubits in the same position in unit cells above and below from the left-hand side and horizontal connections between qubits in the same position in unit cells left and right from the right-hand side of the unit cell.}
\label{fig:chimera}
\end{figure*}

In training via quantum annealing, we map our RBM energy function, which is in the form of a QUBO (quadratic unconstrained binary optimization), and use D-Wave's provided embedding function to map this QUBO into the physical architecture of the D-Wave device, called a Chimera graph, see Fig.~\ref{fig:chimera}.
This is done because the Chimera graph has a maximum complete bipartite subgraph of $4 \times 4$.

By minor embedding \cite{Choi2008MinorembeddingIA} the graph we identify a chain of qubits and bind them tightly together so that they act approximately as a single large spin.
Each programming cycle we use 100 samples drawn from the D-Wave to take our gradient estimate, and apply a varying number (indicated for each experiment) of post-anneal Gibbs sweeps over the variables to aid in additional thermalization.

\section{Boltzmann Distributions on the D-Wave Quantum Annealer}
\label{sec:howtoboltzmann}

In order to train a Boltzmann machine, we need to sample expectation values from a Boltzmann distribution with $\beta$ set to 1, as in~\eqref{eqn:rbmprobdistribution}. 
We use the Kolmogorov-Smirnov (KS) test to check the statistical consistency of our sample distribution with that of a Boltzmann distribution.

\subsection{Temperature estimation}
\label{sec:temp_estimation}

The raw distribution of states coming from a D-Wave 2000Q is often not close to a Boltzmann distribution with $\beta=1$.
It is ``colder'', with a higher propensity for producing states at the lowest energy levels.
This energy shift may be advantageous in optimization problems, but RBM training relies on being able to sample from a Boltzmann distribution, so some correction is generally required.

It is possible that the D-Wave returns a Boltzmann distribution, but at a temperature that needs to be determined.
If we know the effective inverse temperature $\beta_{\text{eff}}$, we can sample from a distribution with $\beta=1$ and couplings $(W, b, c)$ by setting the couplings $J=(W/\beta_{\text{eff}}, b/\beta_{\text{eff}}, c/\beta_{\text{eff}})$ on the D-Wave.
The effective temperature of the D-Wave has been shown to be problem-dependent and different from the physical temperature of the annealer \cite{Amin2015}.
In this work, we will follow a modification of the temperature estimation recipe proposed in Ref.~\cite{Benedetti2016}.

\begin{figure}
    	\centering
		\includegraphics[width=0.99\linewidth]{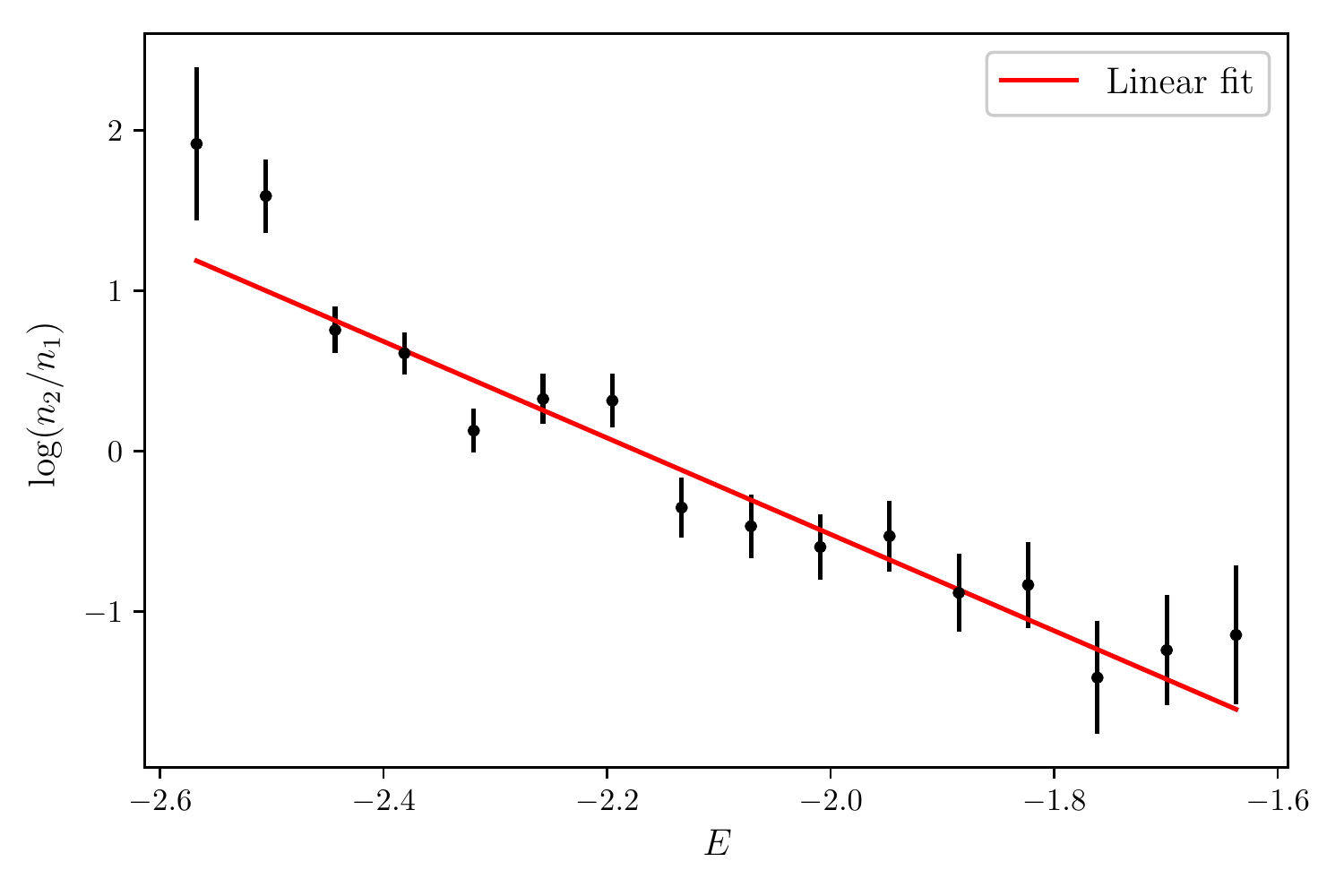}
  		\caption{Linear regression obtained from equation~\eqref{eqn:Testimation_regression} for an example step in training a restricted Boltzmann machine, leading to an estimate of $\beta_{\text{eff}}=3.1$.
  		This regression was obtained at training step 3100 with a batch size of 20, corresponding to epoch 25.}
  		\label{fig:Testimateexample}
\end{figure}

The algorithm follows the following steps:
\begin{enumerate}
\item At each step, take RBM couplings $A=(W, b, c)$.
Set couplings on D-Wave to $J_1=A/\beta_0$, with $\beta_0$ estimated at the previous step (on the first step, we need to take a guess).
\item Take one set of $n$ samples.
We bin the samples into $\left\lceil \sqrt{2n} \right\rceil$ bins according to their energy, obtaining probability density estimates $n_1/n$. \label{firstsample}
\item We want a second set that will provide different ``enough'' samples for distinguishability.
Following \cite{Benedetti2016}, we take $J_2=xJ_1$, with $x=1+1/(\beta_0 \sigma)$,\footnote{\cite{Benedetti2016} suggests transitioning to a $-$ sign in the expression for $x$ once the RBM couplings get large enough.
We found that even at late stages, this would result in values of $x$ that are close to zero.}
where $\sigma$ is the standard deviation of the first sample.
\item Take a second set of samples and use the same bins as in step \ref{firstsample} to obtain probability density estimates $n_2/n$.
\item Denoting the Ising energy of each state with couplings $A$ as $E$, note that
\begin{align}
\frac{n_2}{n_1} &= \frac{e^{-x\beta_{\text{eff}}E/\beta_0}}{Z_2}\frac{Z_1}{e^{-\beta_{\text{eff}}E/\beta_0}} \nonumber \\
\Rightarrow \log \frac{n_2}{n_1} &= \log\frac{Z_1}{Z_2}+(1-x)\frac{\beta_{\text{eff}}}{\beta_0}E. \label{eqn:Testimation_regression}
\end{align}
With this in mind, we can extract an estimate of $\beta_{\text{eff}}$ from the slope of the linear regression between $\log n_2/n_1$ and the bin energies, as exemplified in Fig.~\ref{fig:Testimateexample}.
In order to reduce noise caused by bins with a small number of samples, we limit the regression to bins with at least five samples in both draws.
\end{enumerate}

\begin{figure}
    	\centering
		\includegraphics[width=0.99\linewidth]{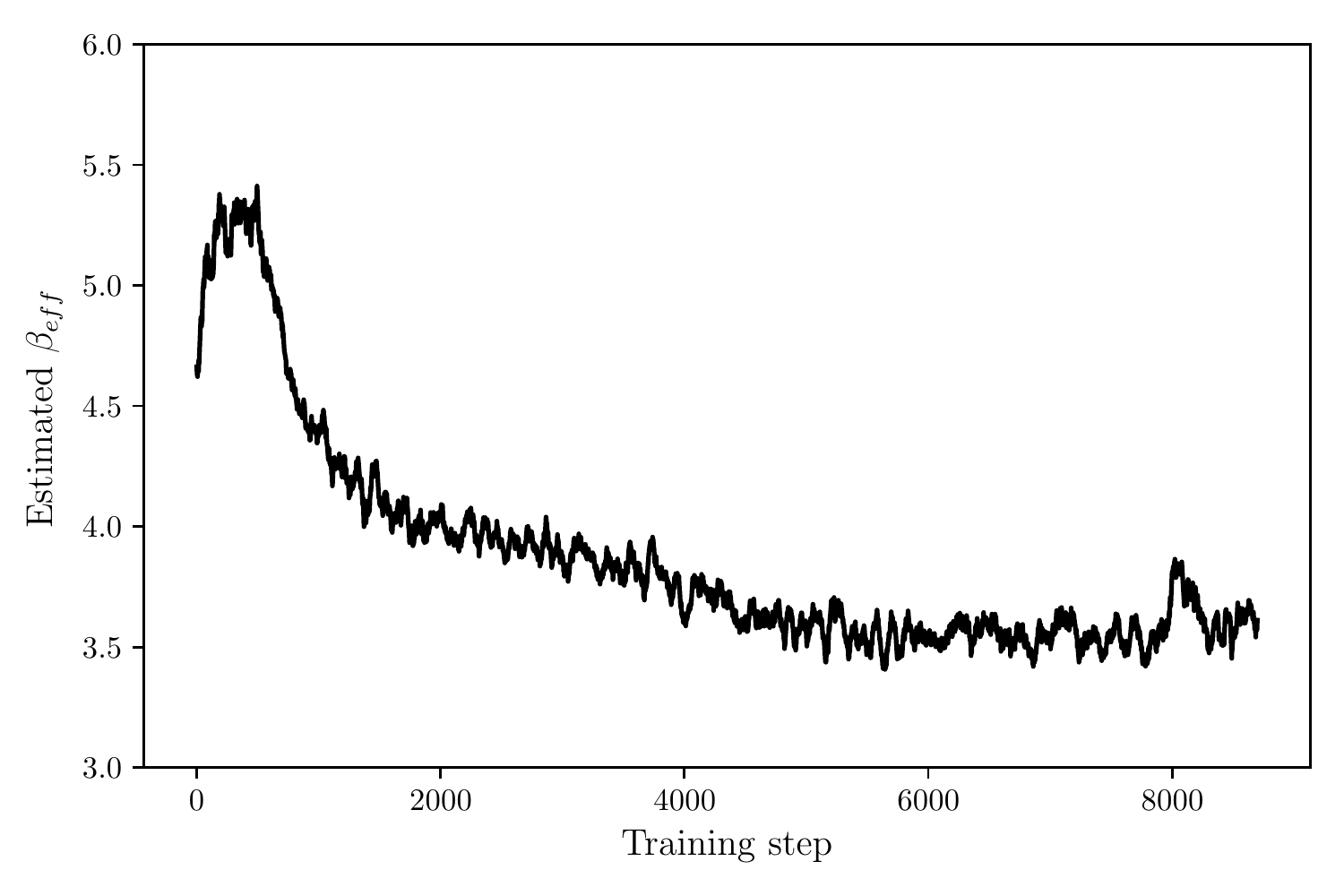}
  		\caption{Temperature estimates over 70 epochs of training (or 8750 training steps) for a $48\times 48$ RBM, plotted using a rolling average over the last 50 steps.
  		It can be seen that the temperature estimates vary significantly over the first stages of training, and later stabilize.
  		This can likely be used to estimate the temperature less often than at every training step.}
  		\label{fig:temperatureestimates}
\end{figure}

We can see the results of this temperature estimation procedure throughout a single training instance in Fig.~\ref{fig:temperatureestimates}.

We found some pitfalls in this procedure.
Namely, as the couplings of the RBM and therefore the magnitude of the energies involved grow, the distribution of states becomes more and more skewed towards the lower energy states.
This is a desirable outcome of training an RBM.
However, this leaves the higher-energy bins with a small number of samples, causing large variance in the estimates of $\log (n_2/n_1)$.
In all our training runs, this leads to a step where $\log (n_2/n_1)$ happens to fluctuate to a larger value than usual for some of the larger energy bins.
This causes $\beta_{\text{eff}}$ to be underestimated at that step.
The effect compounds in a few training steps, often leading to negative estimates of $\beta_{\text{eff}}$ and a crash of the algorithm.

Potential solutions include:
\begin{itemize}
\item some regularization to keep the weights from growing.
This successfully kept the temperature estimation routine from crashing, but at the cost of impairing the classifier performance of our RBM.
This is to be expected, as a well-trained RBM should strongly separate the energies of different states.
\item only estimating $\beta$ during the initial stages of training.
This can be a good solution, since $\beta$ does not seem to change by a large amount during training, as we can see in Fig.~\ref{fig:temperatureestimates}.
\end{itemize}

Even without a temperature estimation routine, weights growing to be too large is a problem with the algorithm on a QA in general.
This is because if weights grow above the maximum coupling that can be implemented on the D-Wave, we must rescale the weights as in order to set coupling constants on the D-Wave.
However, discretization of the coupling constants means that if one weight is very large, subtle variations between much smaller weights are lost.
Another possible solution would be to turn off weight rescaling, but not let couplings grow beyond what is physically implementable on the D-Wave.
Conceptually, this is equivalent to allowing the RBM to learn chains of logical qubits that are strongly coupled.
Either of these solutions can impair classifier performance because sometimes the RBM might just need very large weights, or might need a large ratio across some weights, to reproduce the probability distribution of the data.

As we will see in the following sections, this temperature estimation method does not provide a clear advantage in obtaining Boltzmann distributions for most situations studied here.
Given those results and the pitfalls identified above, we chose to estimate $\beta$ only once or not at all for our longer training runs.

\subsection{Initializing sampling with an annealer vs a random bitstring}
\label{sec:gibbssteps}

Even after temperature estimation, as we will see, the distribution of states returned by the D-Wave will not quite align with the expected Boltzmann distribution.
Therefore, we must use some post-processing.
For us, this will consist of taking a few steps of Gibbs sampling seeded with the D-Wave samples.
When training RBMs, we will use a fixed number of Gibbs sampling steps at each training step.
However, we may ask if that number is enough to reach a Boltzmann distribution.

In this section, to check how many steps are enough to reach a Boltzmann distribution, we carry out the KS test after each Gibbs step and keep taking Gibbs steps until the KS p-value rises above 0.05.
We compare the results of seeding the Gibbs sampling with the samples output by the D-Wave solver to those obtained by seeding it with random bit strings of the same length.
This test is carried out with the couplings set to those of trained 12 by 12 and 48 by 48 RBMs at the end of each training epoch.
We train 20 RBMs for each RBM size and average the results over all networks and a four-epoch window.

\begin{figure}[htp]
\subfloat[Couplings are obtained by training $12 \times 12$ RBMs on the low-noise 2000Q with 10 Gibbs steps.
On this test, there is some advantage to using a D-Wave, especially after estimating $\beta$.]{\label{fig:12_12_stepstoboltzmann}%
  \includegraphics[width=0.48\textwidth]{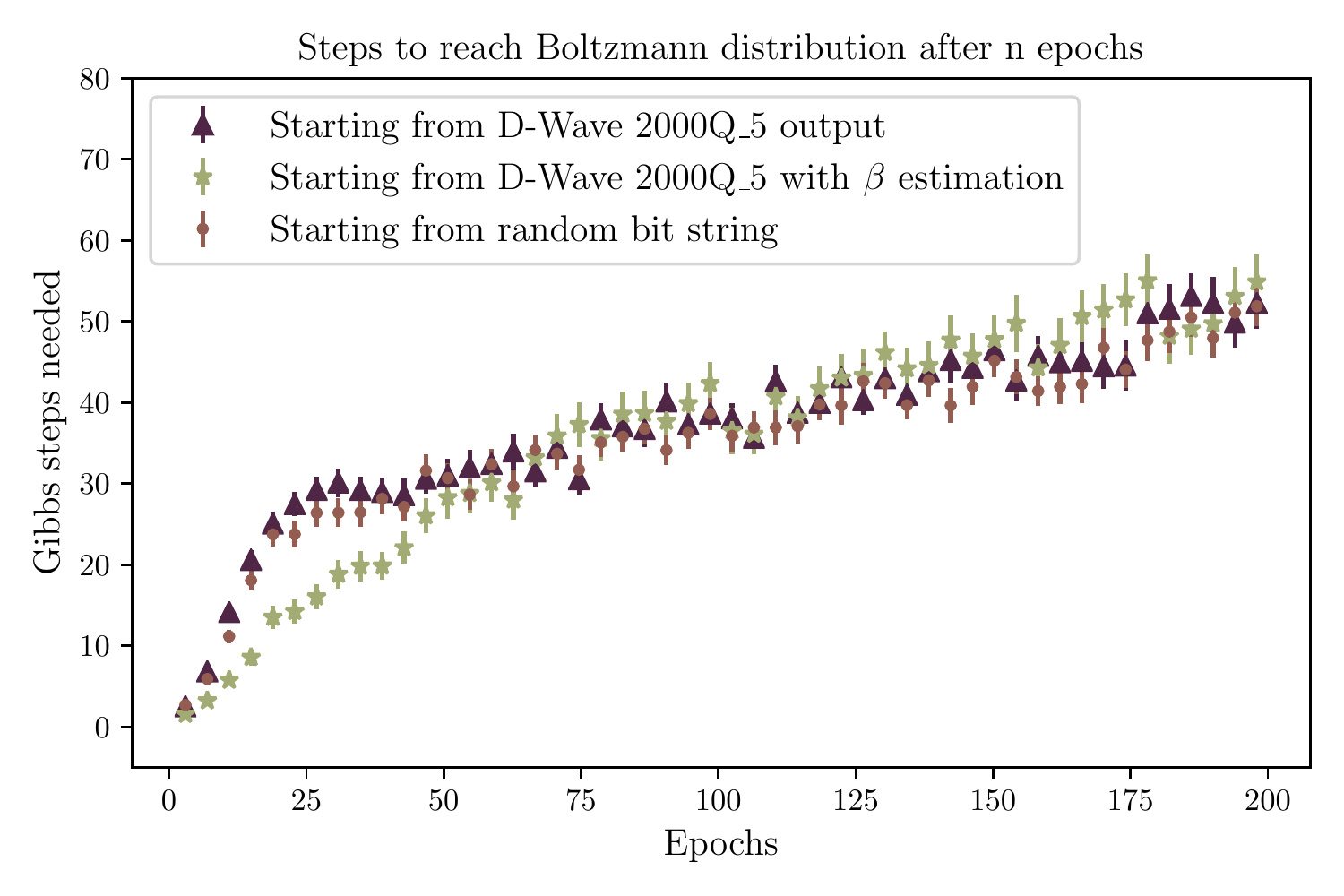}%
} \\
\subfloat[Couplings are obtained by training $48 \times 48$ RBMs on the low-noise 2000Q with 10 Gibbs steps.
In this case, the D-Wave shows some advantage over random strings.]{\label{fig:48_48_stepstoboltzmann}%
  \includegraphics[width=0.48\textwidth]{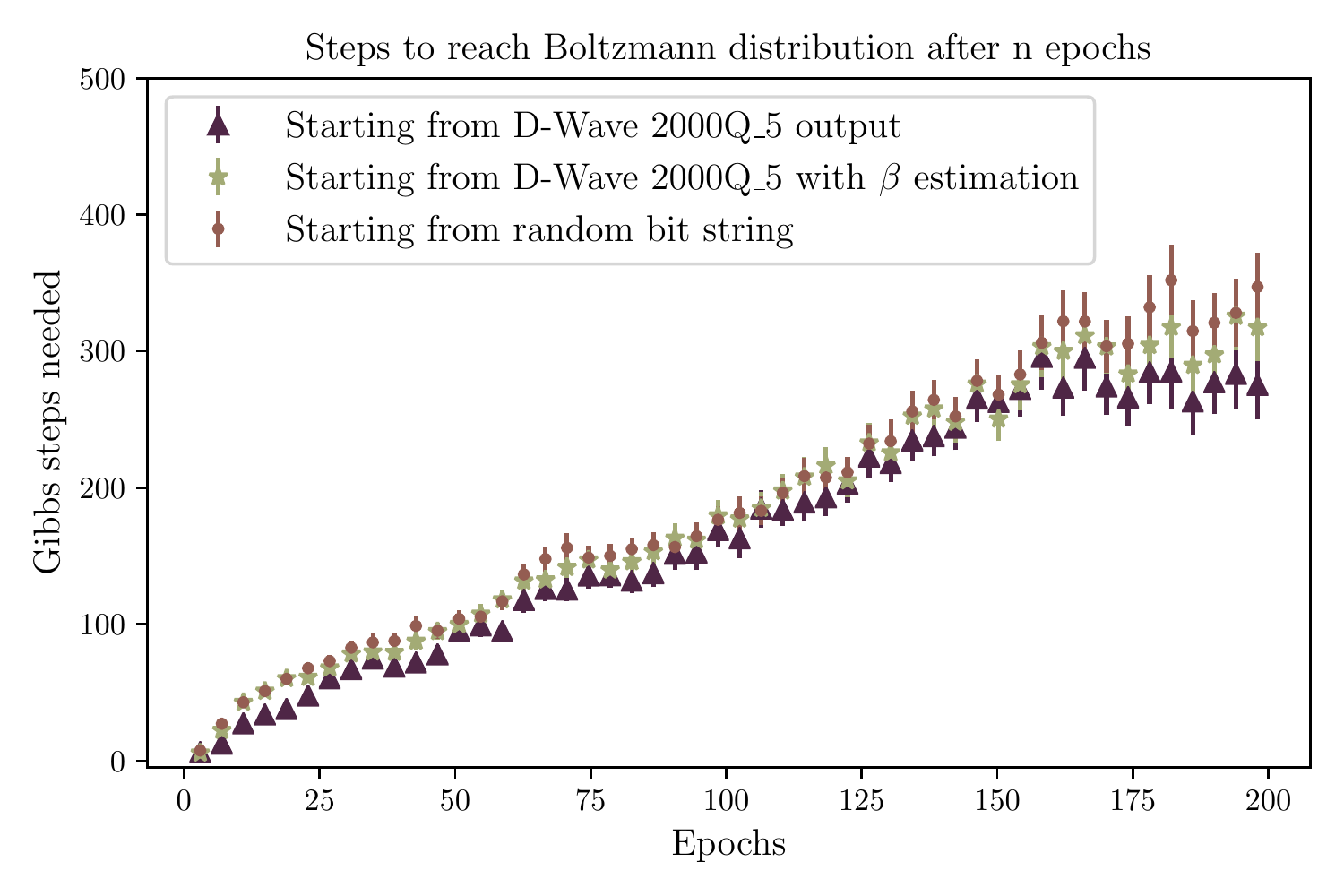}%
}
	\caption{We present results of the test described in section~\ref{sec:gibbssteps}, applied to the D-Wave samples after setting the D-Wave couplings to the actual RBM couplings, and to the RBM couplings scaled by an estimated temperature as in section~\ref{sec:temp_estimation}.
	For the RBM couplings after each training epoch, we apply Gibbs steps until the KS p-value to the true distribution rises above 0.05.
	Note we use the same RBM couplings for all methods in this test, obtained by training using a D-Wave with $\beta$ estimated only at the first training step and used throughout training, with batch size set to 20 and 10 Gibbs steps taken as post-processing after sampling.
	The figure uses 50 bins, averaging the results of four epochs per bin for twenty independently trained RBMs.
	Error bars show the standard error on the mean.
	}
	\label{fig:stepstoboltz}
\end{figure}

\begin{figure*}[htp]
\subfloat[Ratio of the means presented in \ref{fig:12_12_stepstoboltzmann} for 12 by 12 RBM. It is clear from this that $\beta$ estimation is beneficial for early epochs, but the advantage is lost after about 50 epochs.]{\label{fig:12_12_steps_ratio}%
  \includegraphics[width=0.48\textwidth]{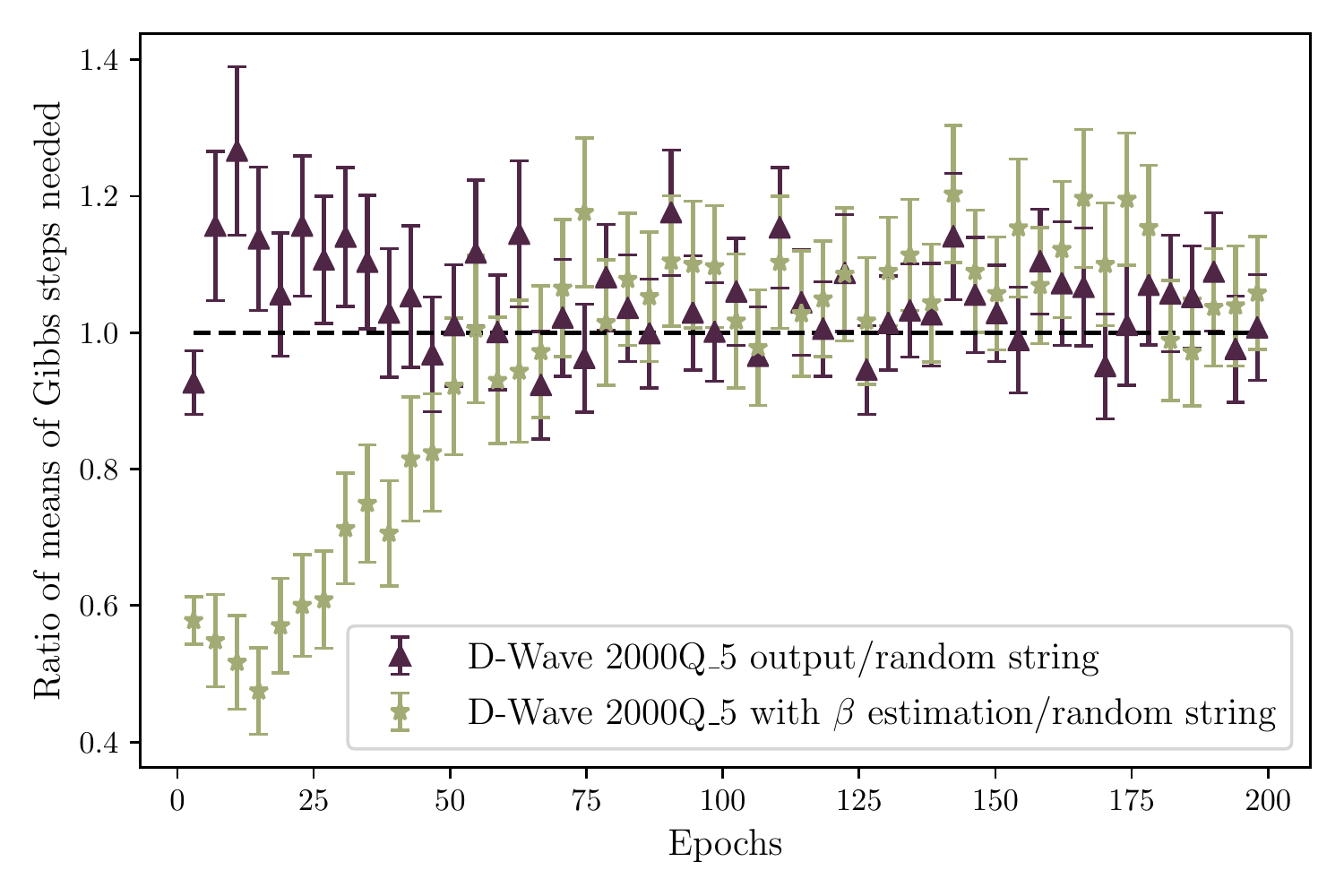}%
} \hfill
\subfloat[Fitting a Bernoulli random variable to these distributions confirms the advantage of $\beta$ estimation for the first 50 epochs in smaller RBM (here, 12 by 12).]{\label{fig:12_12_bernoulli}%
  \includegraphics[width=0.48\textwidth]{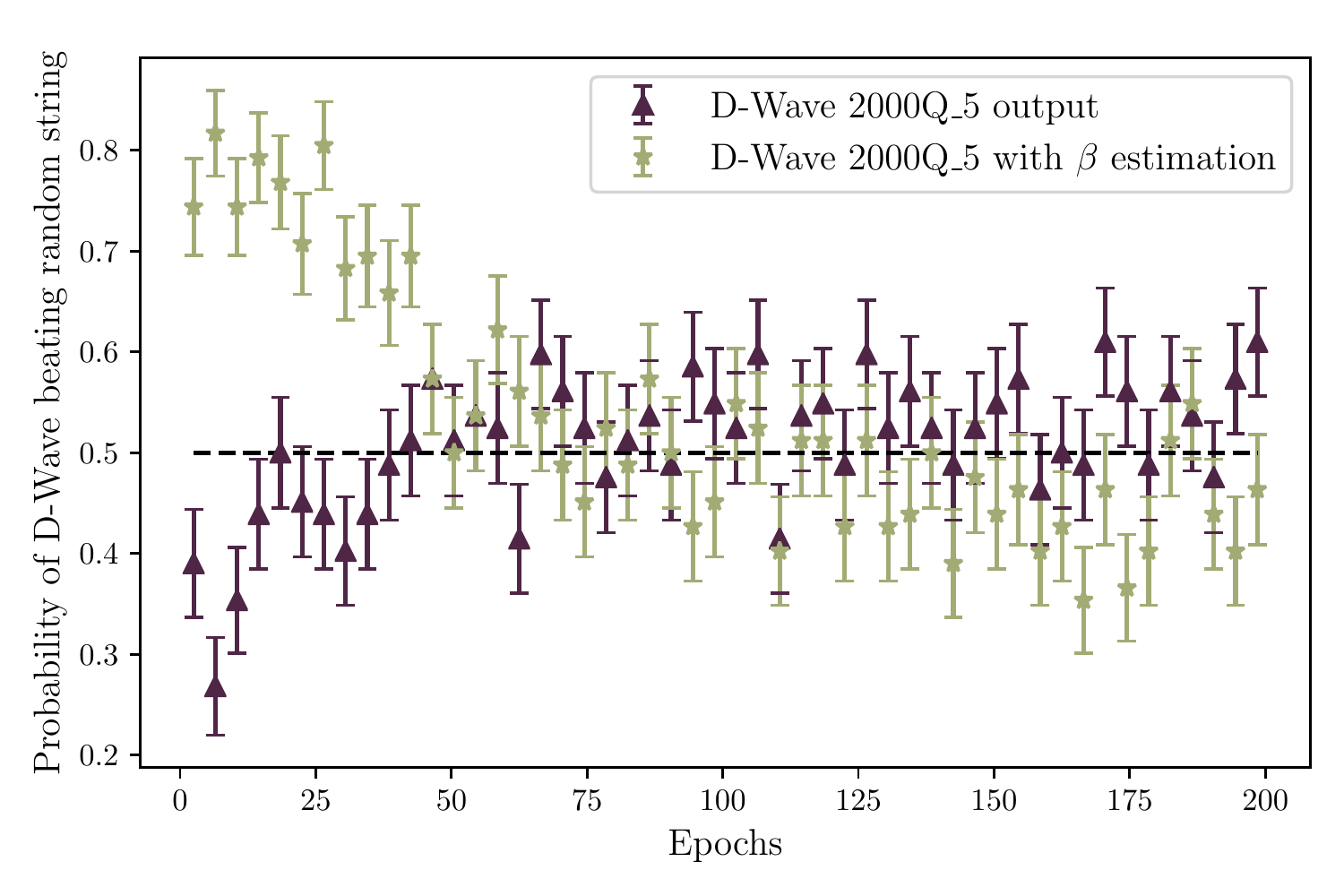}%
} \\
\subfloat[Ratio of the means presented in \ref{fig:48_48_stepstoboltzmann} for 48 by 48 RBM. Sampling from a D-Wave leads to a smaller number of mean Gibbs steps needed throughout all epochs of training.]{\label{fig:48_48_steps_ratio}%
  \includegraphics[width=0.48\textwidth]{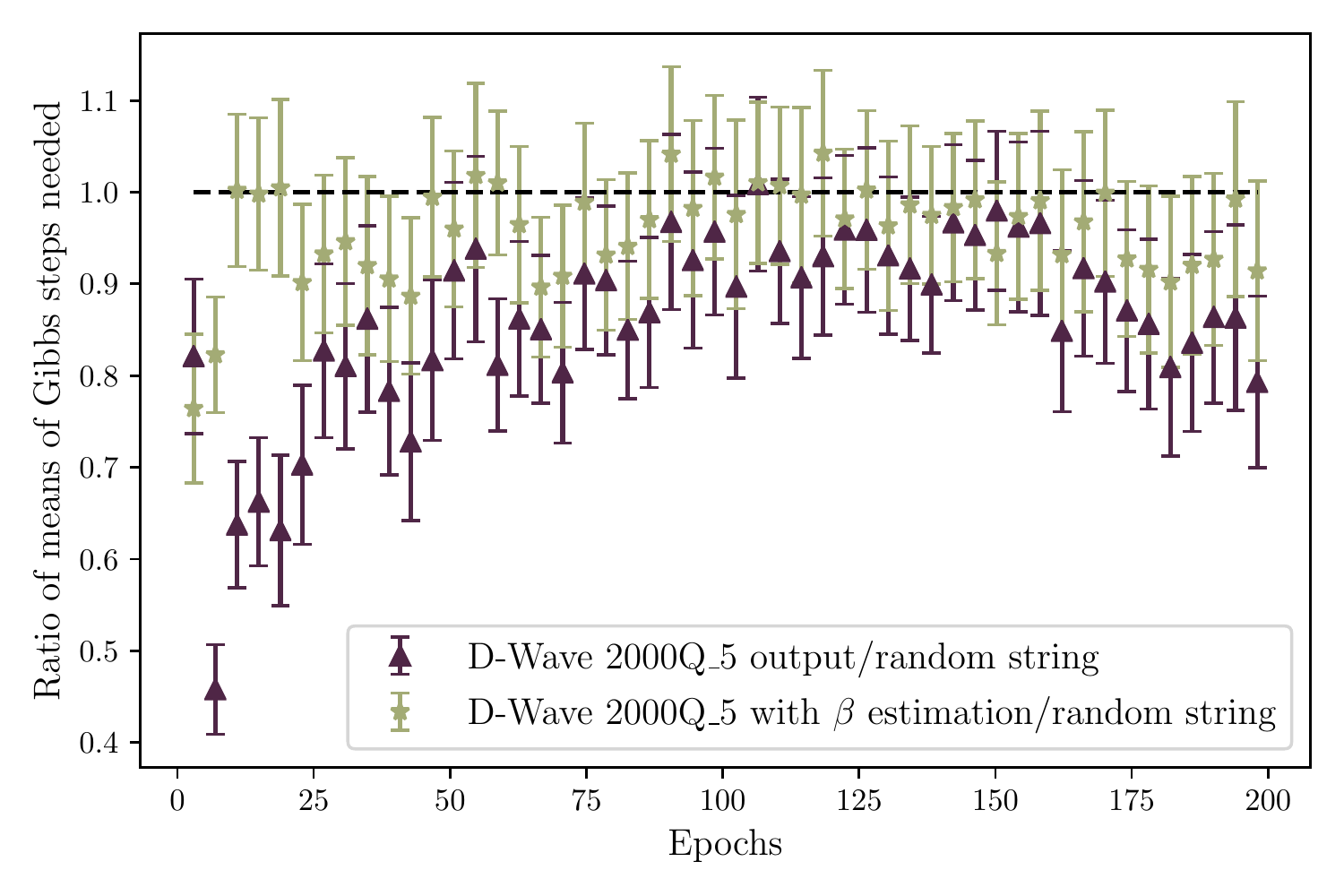}%
} \hfill
\subfloat[For larger RBM (48 by 48), while sampling from a D-Wave leads to a smaller number of average Gibbs steps needed, we are only likely to need a smaller number of steps in early epochs.]{\label{fig:48_48_bernoulli}%
  \includegraphics[width=0.48\textwidth]{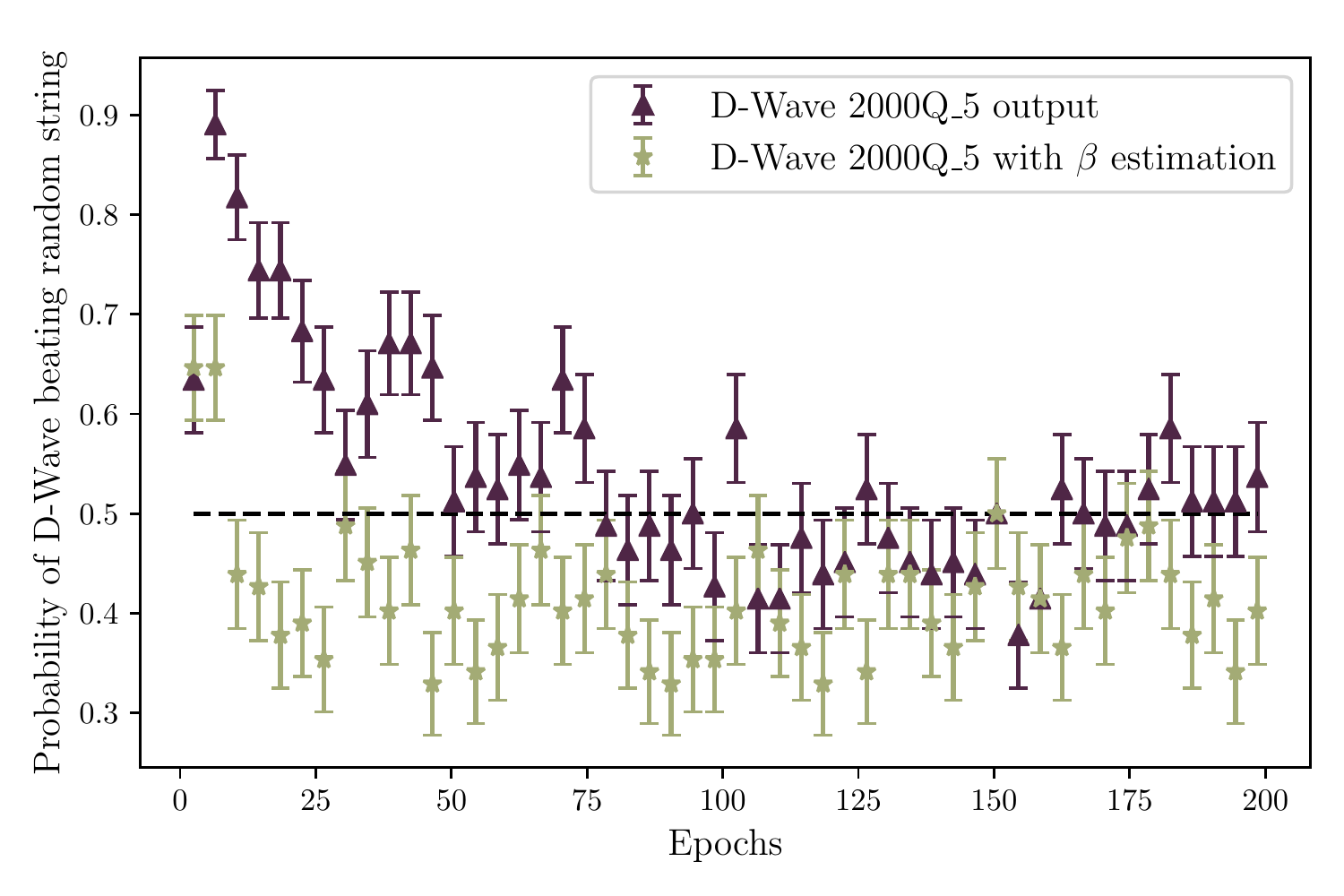}%
}
	\caption{We present more results of the test described in section~\ref{sec:gibbssteps}.
	On the left, we show the ratios of the means presented in Fig.~\ref{fig:stepstoboltz} to help comparisons.
	We indicate a ratio of 1, corresponding to equivalent methods.
	On the right, we fit a Bernoulli distribution to the random variable indicating whether we need fewer Gibbs steps to reach a Boltzmann distribution when starting from a D-Wave sample.
	As in Fig.~\ref{fig:stepstoboltz}, we use the same RBM couplings for all methods.
	}
	\label{fig:stepstoboltz_processed}
\end{figure*}

The mean number of steps needed at each training stage can be seen in Fig.~\ref{fig:stepstoboltz}.
To ease the comparison of the averages, we take their ratio in Fig.~\ref{fig:stepstoboltz_processed}.
Finally, since averages can be affected by outliers, we also present in Fig.~\ref{fig:stepstoboltz_processed} the results of fitting a Bernoulli distribution to the random variable representing needing fewer Gibbs steps after starting from the D-wave samples than when starting from a random string.
For RBMs with 12 hidden and 12 visible units, the advantage exists for samples taken after scaling by the estimated $\beta$ for early epochs.
For later epochs and for results obtained without temperature estimation, we see no advantage in this test.
For larger RBM, namely $48 \times 48$, $\beta$ estimation seems to become less important.
However, samples taken without $\beta$ estimation take fewer Gibbs steps to reach a Boltzmann distribution, showing an advantage over starting with a random bit string across all epochs, especially early in training.

\begin{figure}[htp]
\subfloat[12 by 12 RBM steps as a function of median coupling.]{\label{fig:12_12_stepsbycoupling}%
  \includegraphics[width=0.49\linewidth]{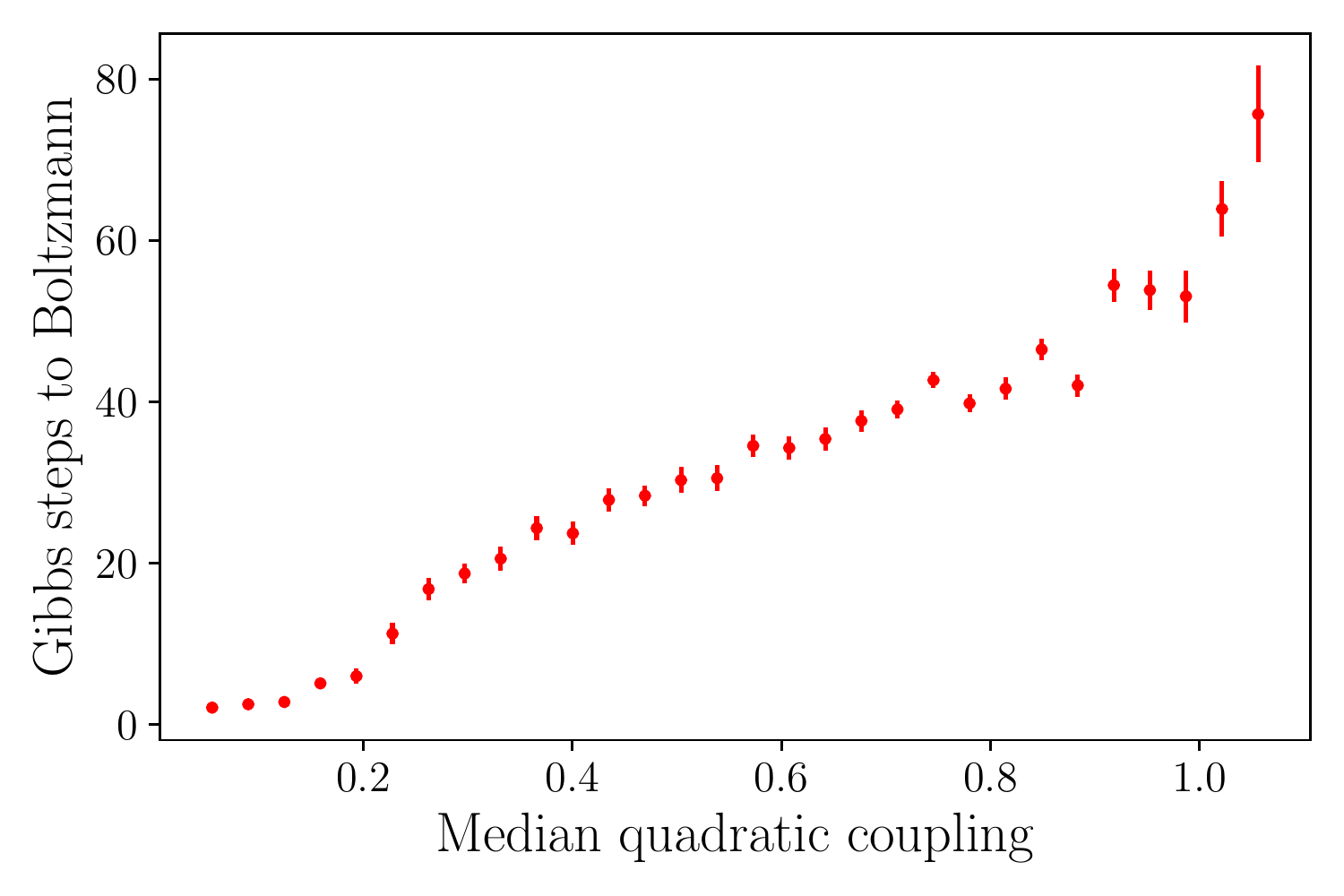}%
} \hfill
\subfloat[12 by 12 RBM median coupling as a function of epoch.]{\label{fig:12_12_couplingperepoch}%
  \includegraphics[width=0.49\linewidth]{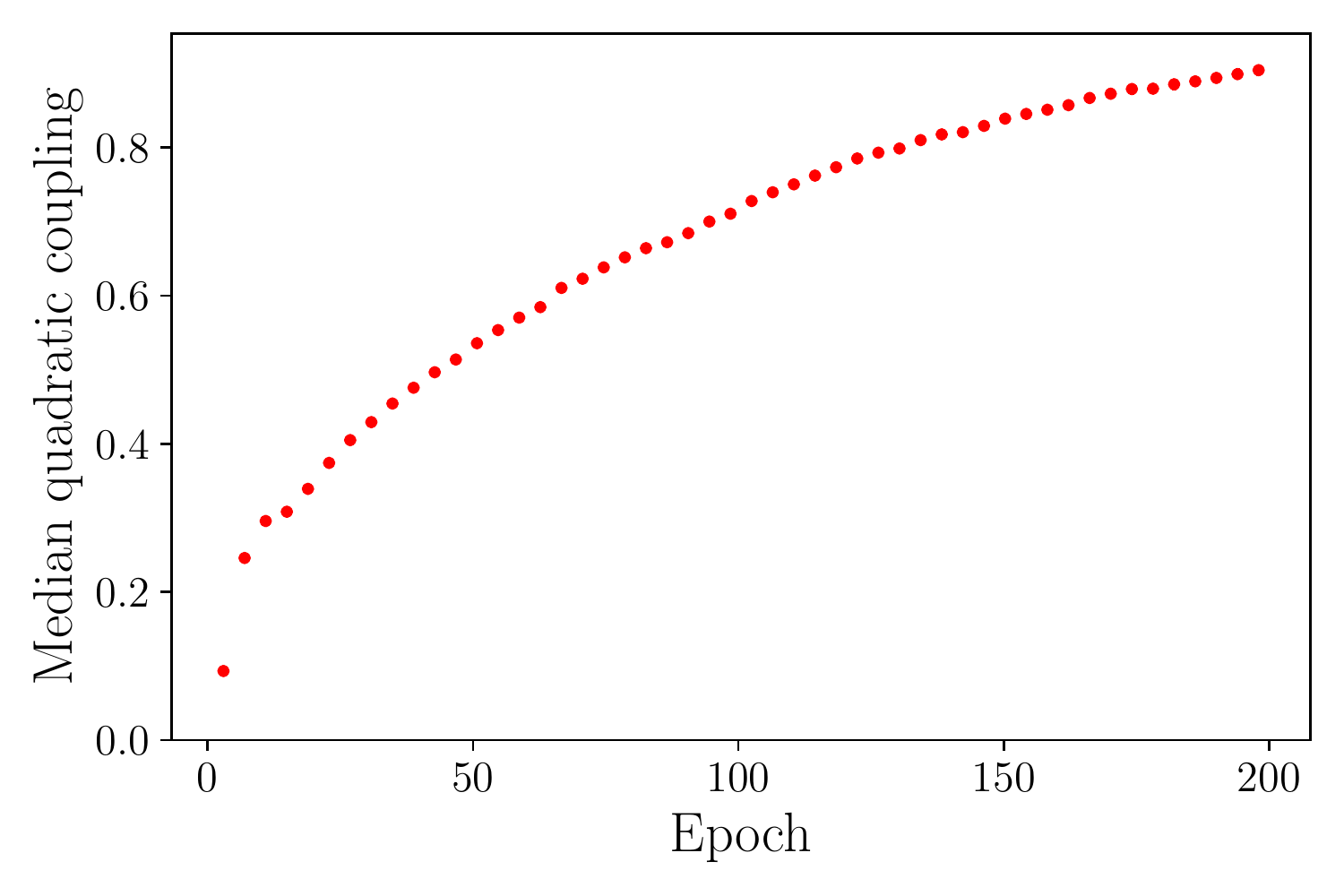}%
}
\\
\subfloat[48 by 48 RBM steps as a function of median coupling.]{\label{fig:48_48_stepsbycoupling}%
  \includegraphics[width=0.49\linewidth]{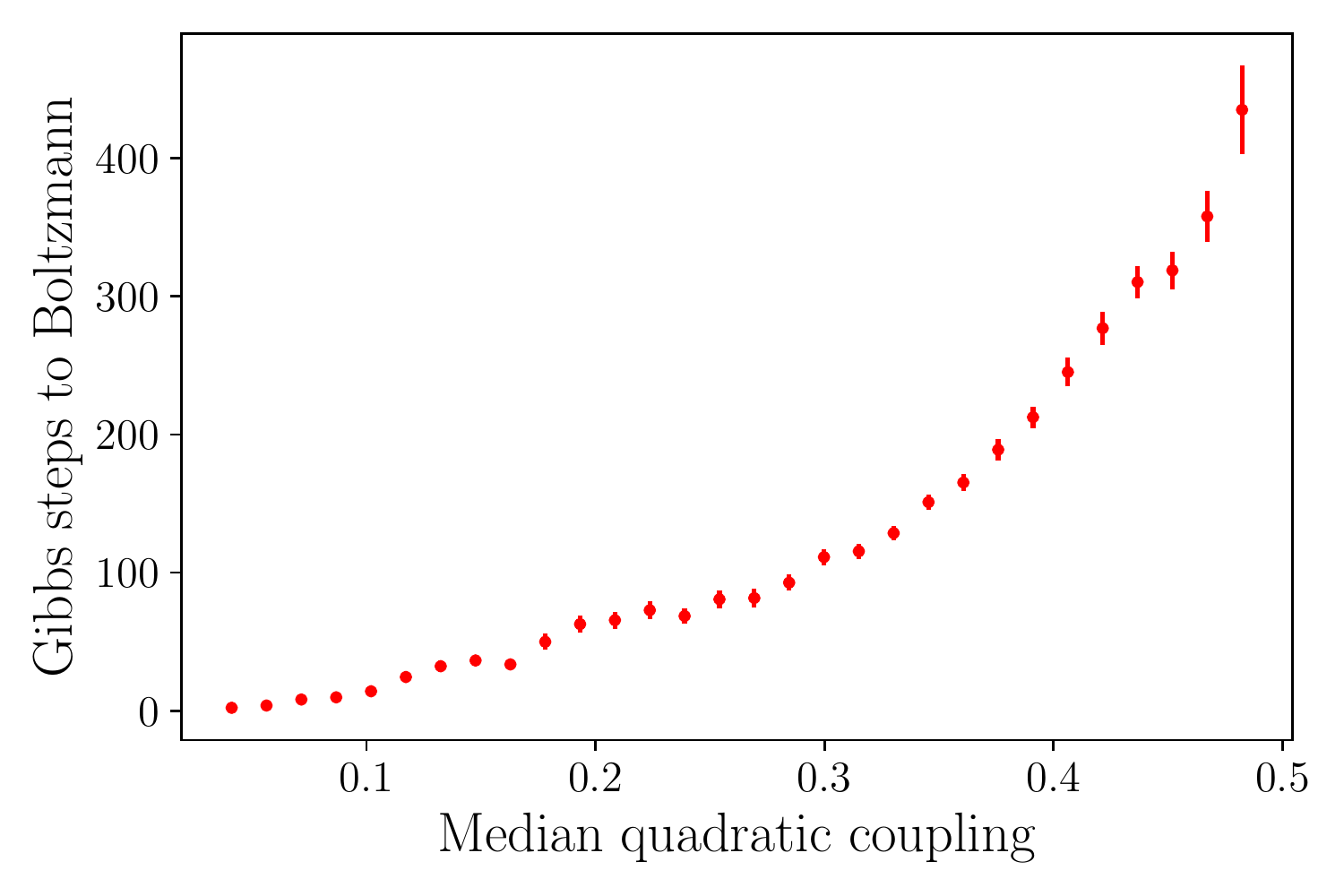}%
} \hfill
\subfloat[48 by 48 RBM median coupling as a function of epoch.]{\label{fig:48_48_couplingperepoch}%
  \includegraphics[width=0.49\linewidth]{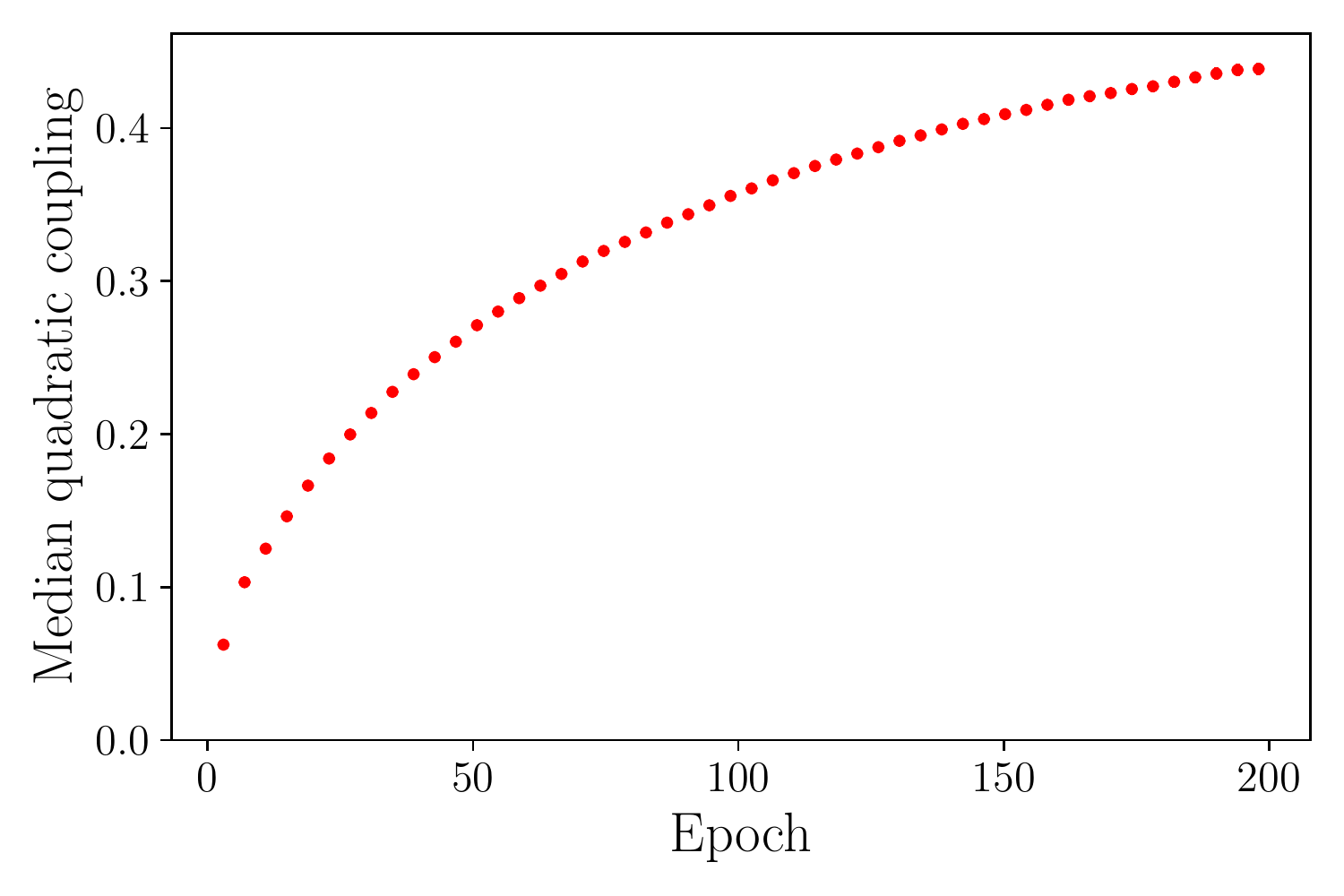}%
}
	\caption{Gibbs steps as a function of the median coupling, and median RBM quadratic coupling as a function of epoch.
	The faster-than-linear evolution of the number of Gibbs steps needed for different coupling values is approximately canceled by the slower-than-linear change in the couplings as a function of epoch.
	This leads to the approximately linear behavior in Fig.~\ref{fig:stepstoboltz}.
	}
	\label{fig:stepsbycoupling}
\end{figure}

One interesting observation is that the number of Gibbs steps needed to reach a Boltzmann distribution grows approximately linearly with the epoch number.
To understand this better, we can see in Fig.~\ref{fig:stepsbycoupling} that how the number of steps grows with the median quadratic coupling of the RBM, and how the median quadratic coupling grows with epoch.
These two non-linear trends seem to conspire leading to the approximately linear growth in Fig.~\ref{fig:stepstoboltz}.

\subsection{Noise and RBMs}
\label{sec:noise}

D-Wave has recently released a low-noise version of its 2000Q quantum computer, with claims to enhancing tunneling rates by a factor of 7.4 \cite{qubits_pres}.
It is claimed that this leads to a larger diversity of states returned by the machine, as well as a larger proportion of lower-energy states.
In this section, we test whether these lower-noise properties also help us obtain a more Boltzmann-like distribution from the D-Wave output.
To do this, we train two $12 \times 12$ RBM using the temperature estimation techniques described in Sec.~\ref{sec:temp_estimation}.
One of the RBM was trained using the original 2000Q, and the other RBM using the low-noise machine.
At each 20 training steps, we compare the distribution obtained using the D-Wave machine with a Boltzmann distribution obtained from analytically calculated energies for the current RBM couplings.
To compare the distributions, we use the Kolmogorov-Smirnov statistic, which should be close to zero for samples drawn from the same distribution.
We compare the KS values as a function of the RBM weight distribution in each machine.

\begin{figure}
\subfloat[KS statistic as a function of maximum RBM coupling after scaling by the effective $\beta$.]{\label{fig:maxwvsks}%
  \includegraphics[width=0.99\linewidth]{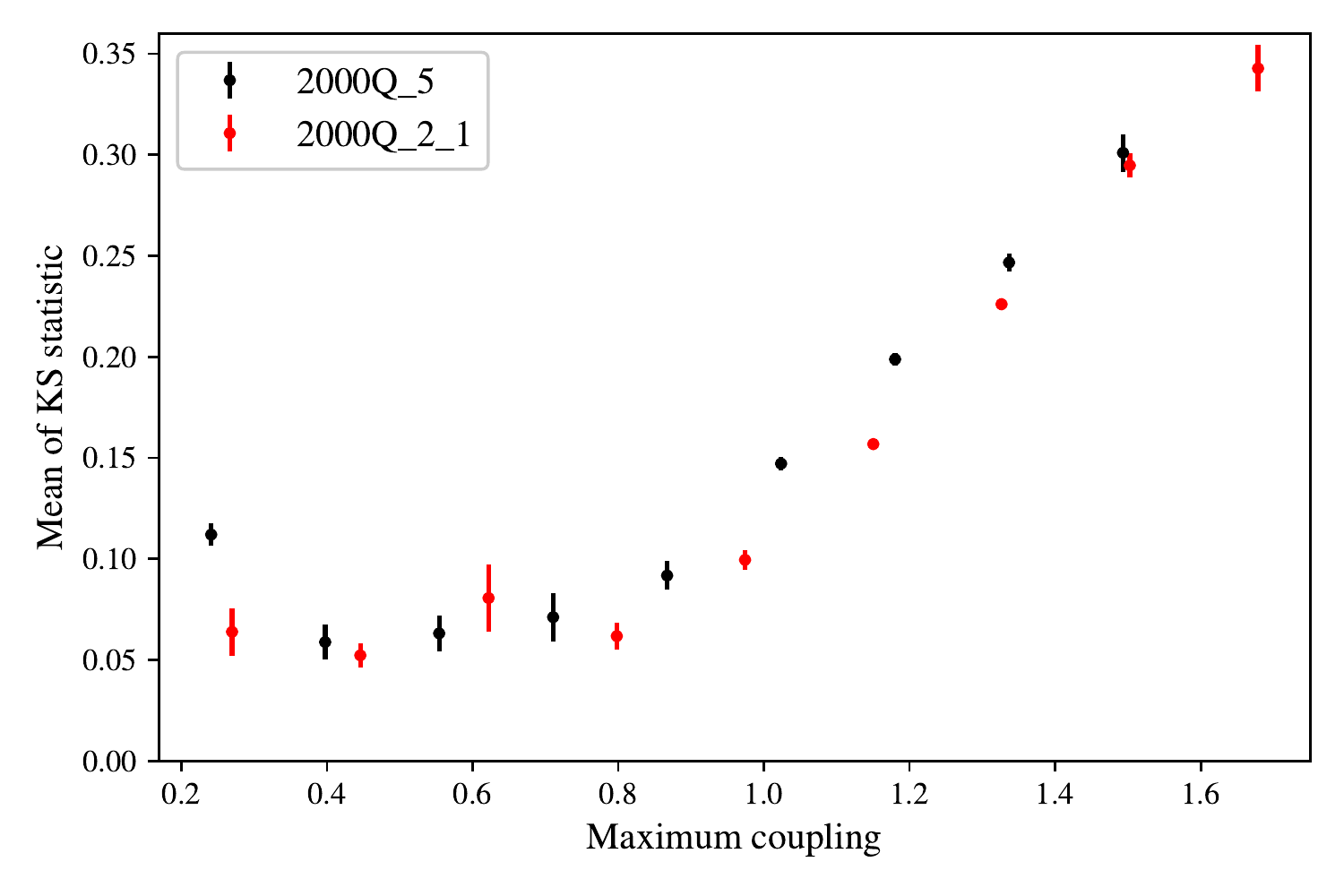}%
} \\
\subfloat[KS statistic as a function of mean RBM coupling after scaling by the effective $\beta$.]{\label{fig:meanwvsks}%
  \includegraphics[width=0.99\linewidth]{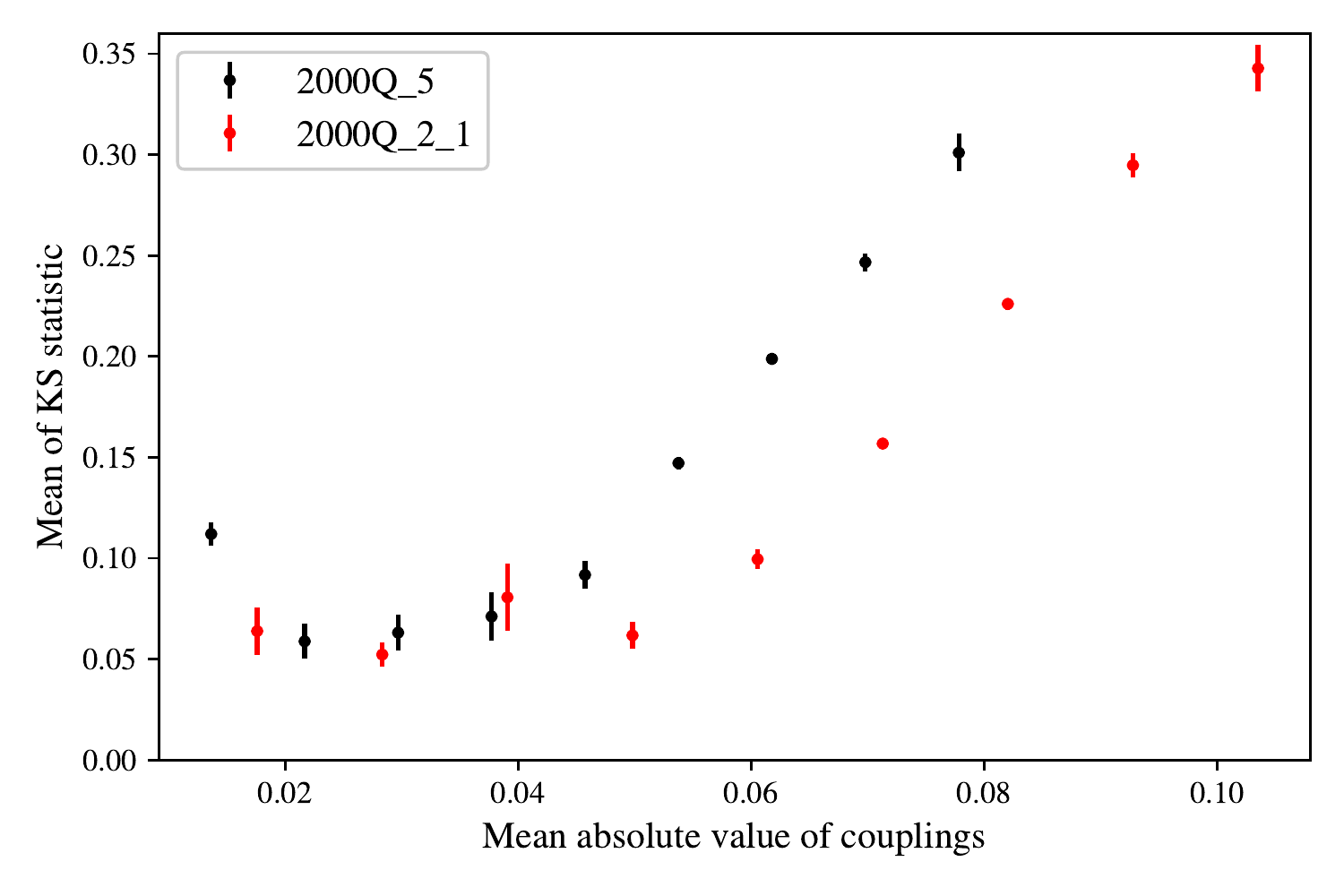}%
}
	\caption{We compare the KS statistics between Boltzmann samples and D-Wave samples for two 12 by 12 RBM, one trained on the original 2000Q and one trained on the low-noise 2000Q.
	To compute the KS statistics, we compare 1000 samples from the D-Wave to 1000 random samples taken from the correct Boltzmann distribution.
	We see no advantage in finding a Boltzmann distribution from using the low-noise 2000Q.
	Note each machine was tested using couplings of an RBM trained on that same machine, so the range of tested couplings differs slightly.
	}
	\label{fig:lownoisecomp}
\end{figure}

In Fig.~\ref{fig:lownoisecomp}, we show the mean of the KS statistic binned as a function of the mean and maximum RBM coupling.
We see no advantage from using the lower-noise 2000Q in how Boltzmann-like the returned distributions are.
For both machines, samples returned are not far from Boltzmann distributions (with KS statistics below 0.1) for low RBM weights, but the distributions diverge from Boltzmann as the weights grow larger.

\begin{figure}
    	\centering
		\includegraphics[width=0.99\linewidth]{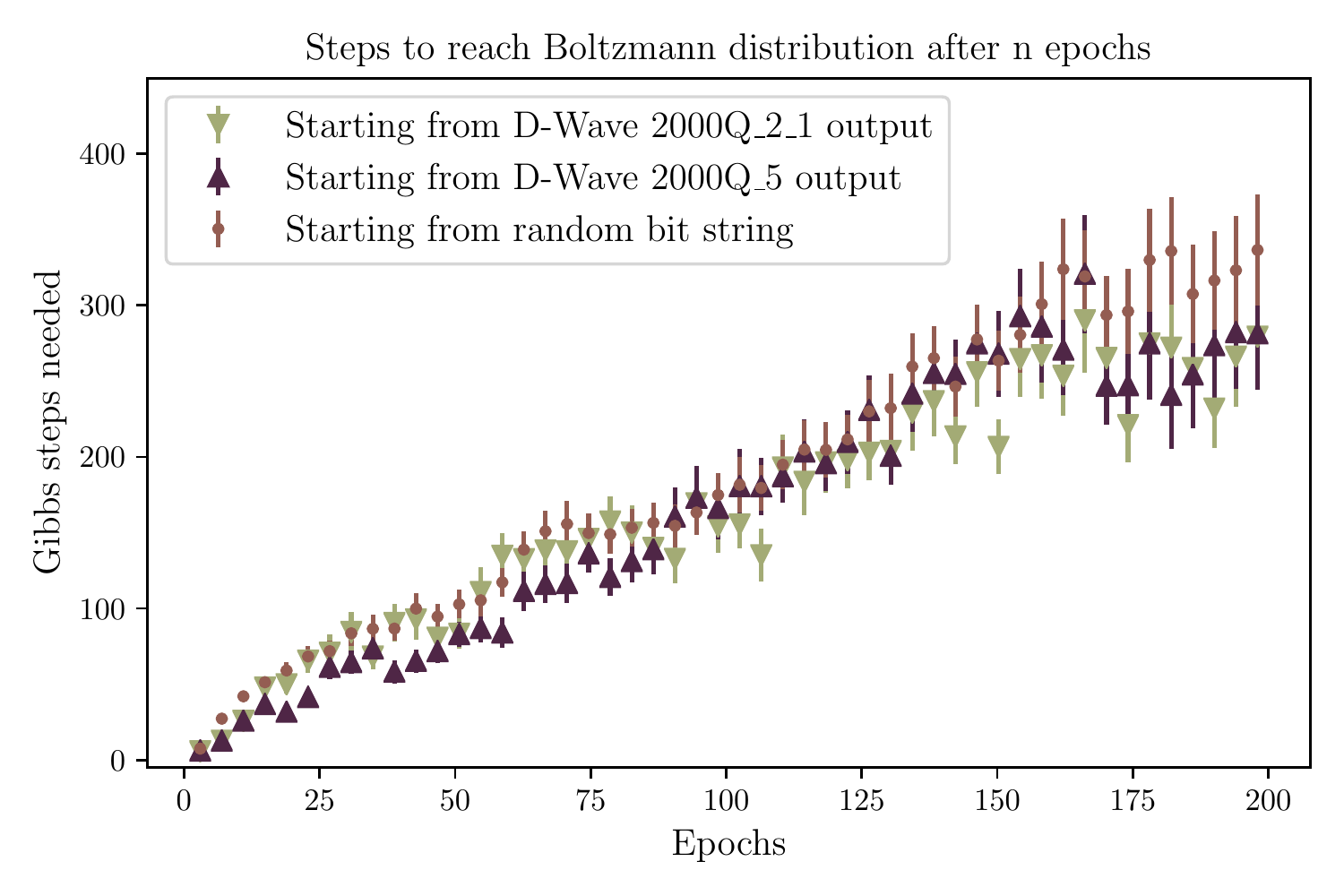}
  		\caption{We present the results of a comparison similar to that of Fig~\ref{fig:stepstoboltz}, but between obtaining samples on the older 2000Q\_2\_1 machine compared to the current low-noise 2000Q\_5 annealer.
  		We see no significant difference between using the older 2000Q and the low-noise 2000Q.
  		As in Fig.~\ref{fig:stepstoboltz}, we use the same RBM couplings for both methods.} \label{fig:stepstoboltz_lownoise}
\end{figure}

We also try a test similar to Fig.~\ref{fig:stepstoboltz}, initializing the Gibbs steps with samples from either 2000Q machine for RBM couplings set to those occurring at the end of epochs in ten RBM training runs.
The results are shown in Fig.~\ref{fig:stepstoboltz_lownoise}.

\section{Results and discussion}
\label{sec:results}

Here we focus on comparing training accuracy, defined as the total fraction of all testing data correctly classified by the trained RBMs, between the wide variety of algorithms discussed in section \ref{sec:generativediscriminative}, along with straightforward logistic regression and gradient boosted decision trees, as a function of training epoch. For results given in this section, we use two Gibbs sweeps of post-processing for quantum annealing samples, except as otherwise stated.

\begin{figure*}[hp]
\subfloat[Comparison of algorithm accuracy for standard classification techniques, logistic regression and gradient boosted decision trees, against quantum-based RBM training along with RBM training using efficient discriminative RBM training for different sized RBMs.
The classical models strictly outperform all RBM-basd models. Note: epoch for the tree-based model corresponds to the number of decisions trees.]{\label{fig:bigfreakingplot}%
  \includegraphics[width=0.99\textwidth]{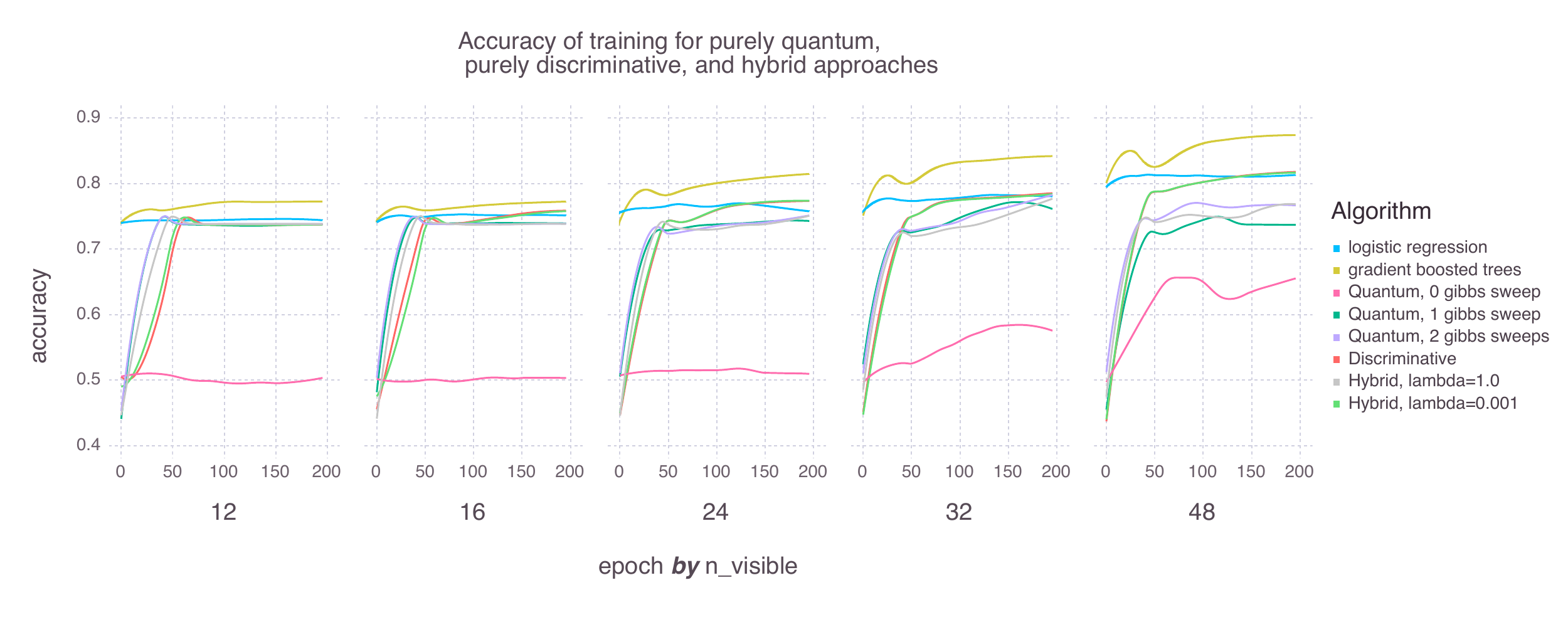}%
} \\
\subfloat[Comparison of various algorithms accuracy as a function of the number of epochs for difference sized RBM-structured models.
In summary, quantum training appears to broadly achieve higher accuracy early in training but either match (for small RBMs) or underperform relative to pure discriminative and hybrid training with small values of $\lambda$ (for large RBMs).
Transitioning from QA to discriminative training at approximately the localtion of the crossover in performance does not yield as high an accuracy at the end of 100 epochs as pure discriminative training.
Data shown is for batch size of 128.
Moreover, QA-style results seem to be approximately reproducible with fairly brief simulated annealing runs.
The best final performance is from purely discriminative training.]{\label{fig:bigfreakingplot_genvdisc}%
  \includegraphics[width=0.99\textwidth]{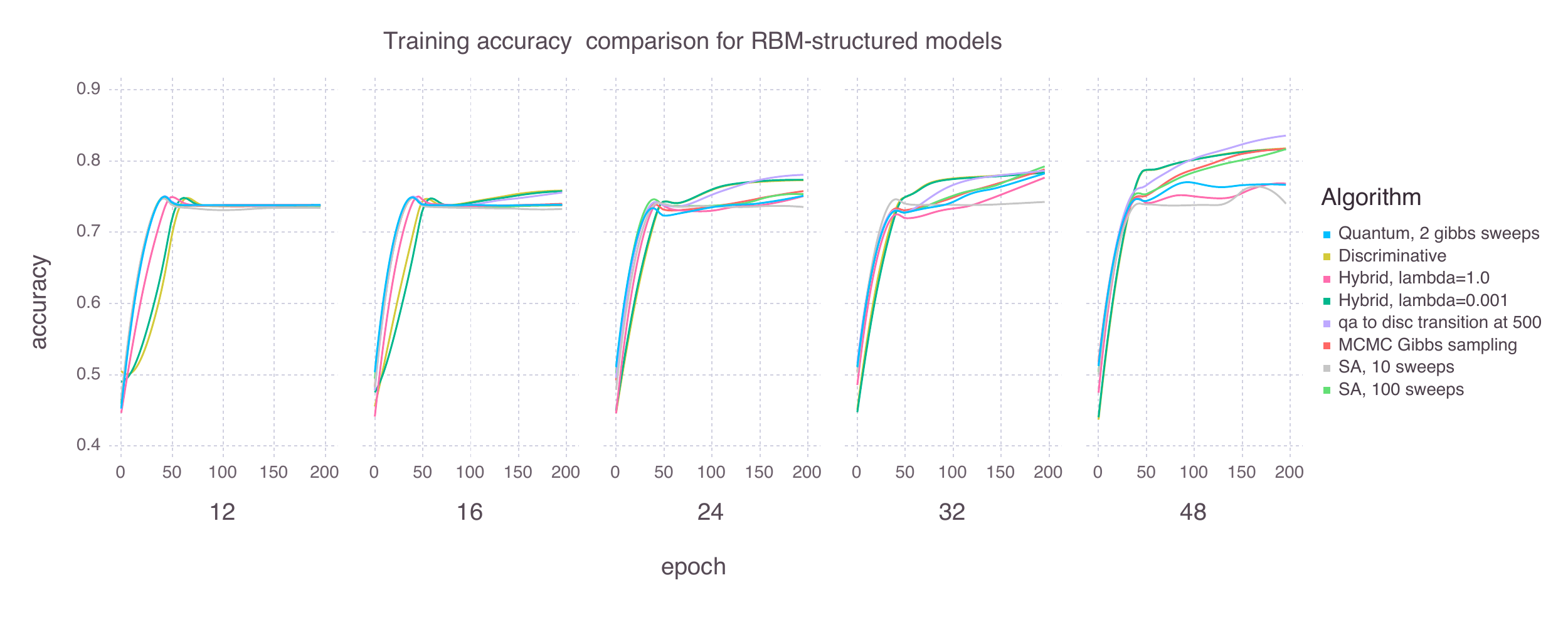}%
}
\caption{}
\label{fig:thebigone}
\end{figure*}

We present a comprehensive figure of our primary results in Fig.~\ref{fig:thebigone}. 
In Fig.~\ref{fig:bigfreakingplot} we see that QA requires at least two Gibbs sweeps to perform competitively.
Moreover there we also see, all RBM-structured models (which are upper-bounded, as seen in Fig.~\ref{fig:bigfreakingplot_genvdisc}, by discriminative training) are outperformed for this problem by simple logistic regression and particularly gradient boosted decision trees. (Note: ``epoch'' for gradient boosted decision trees corresponds to the number of trees in these plots.)
Looking at Fig.~\ref{fig:bigfreakingplot_genvdisc} (where we focus on RBM-structured models),  we show that while QA appears to achieve higher accuracy at early stages of training than other RBM-structured models/training methods, as training progresses it either matches (for smaller size RBMs) or underperforms other algorithms such as purely discriminative and hybrid training.
Moreover, it appears to lend little if anything to discriminative training to incorporate hybrid updates using QA, or to transition from QA to discriminative training near the observed crossover point of performance.
Unless one is only going to run training for a short period (on the order of 25 epochs), one observes no improvement from the incorporation of QA.
We also examine performance of MCMC Gibbs sampling (ie directly taking expectation values from a Markov chain of appropriate length) and simulated annealing, and observe broadly similar performance as QA, with small improvements for MCMC and SA over QA at larger size RBMs.

\begin{figure*}[hp]
\includegraphics[width=0.99\textwidth]{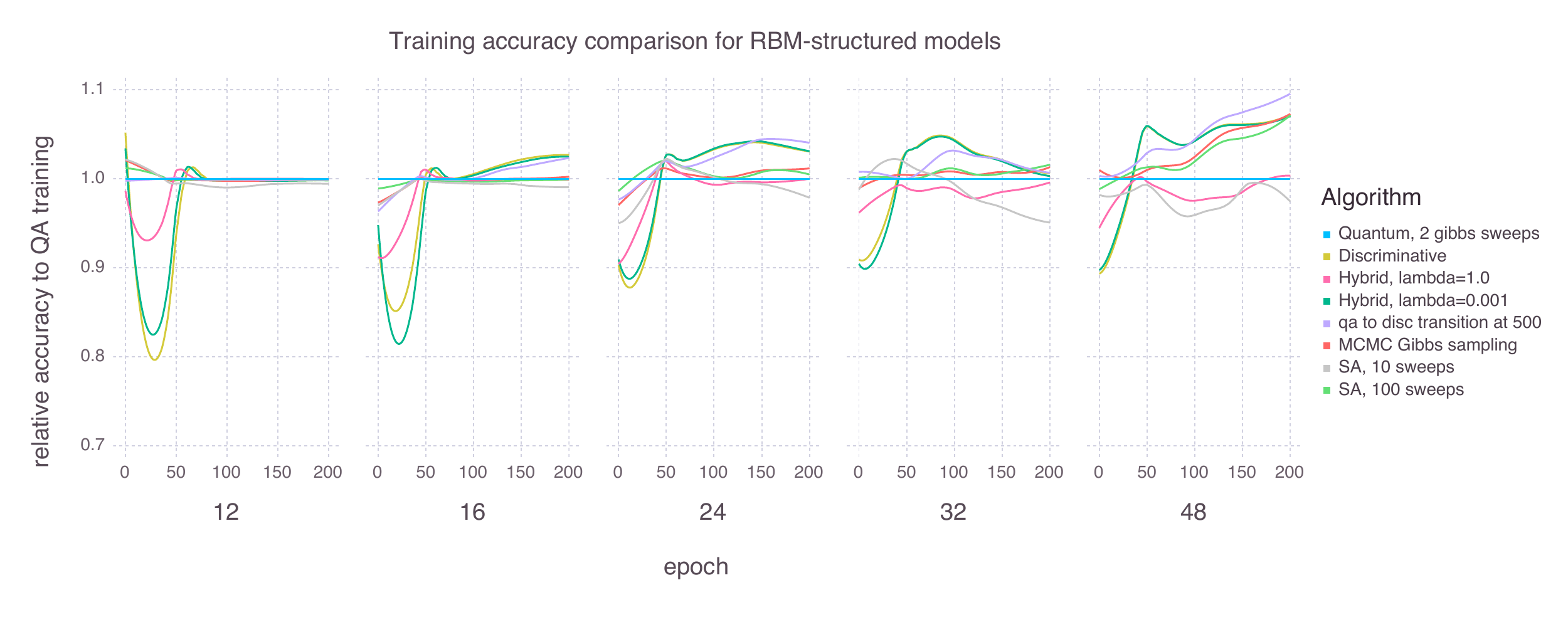}%
\caption{Comparison of the ratio of various algorithms accuracy against the accuracy of QA training (with 2 Gibbs sweeps) per epoch.
Values larger than one imply higher accuracy for the algorithm than QA, below one worse accuracy. This only displays RBM-structured models.
Logistic regression and gradient boosted trees dominate RBM-structured models and are not included so as to maximize resulution in the comparison among RBM-structured models. Note: epoch for the tree-based model corresponds to the number of decisions trees.}
\label{fig:algcomp_ratio}
\end{figure*}

We also present Fig.~\ref{fig:algcomp_ratio} to highlight in more detail the relative performance of the various RBM-structured algorithms, wherein we take the ratio of their accuracy at each step of training with the accuracy of the quantum annealing training.
As this figure makes clear, QA achieves better early training results but fails to maintain that advantage with additional training. It also makes clear that MCMC Gibbs sampling and SA with sufficiently many sweeps outperforms QA slightly at larger RBM sizes. In that figure we exclude logistic regression and gradient boosted decision trees as they dominate all models.

\begin{figure*}[hp]
\includegraphics[width=0.99\textwidth]{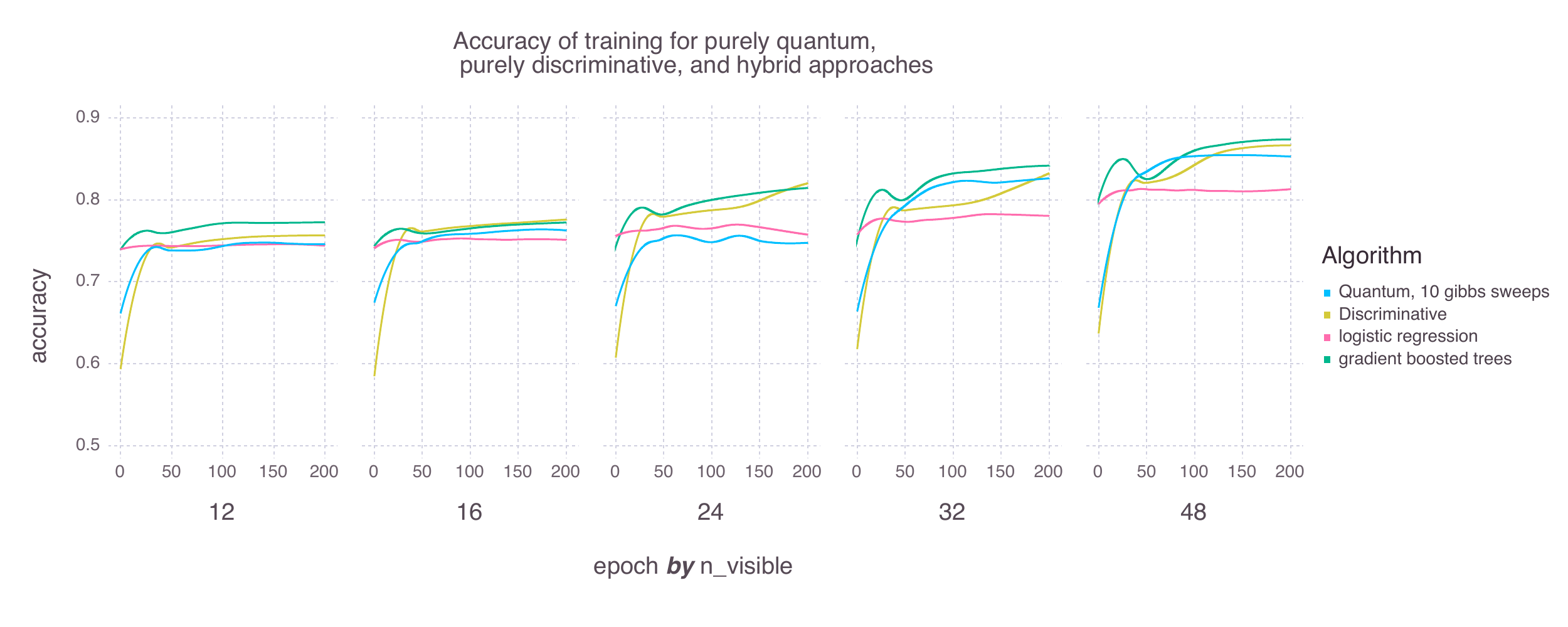}
\caption{Small batch size comparison of algorithm accuracy for logistic regression and gradient boosted decision trees against quantum-based and efficient discriminative training for different sized RBMs. The RBM models were trained with a batch size of 20, as opposed to 128 for previous performance comparisons. Performance for the RBM models improves compared to larger batch sizes, and we observe some short scale improvement of QA over discriminative training. However by 200 training epochs discriminative training again outperforms QA and both are uniformly outperformed by gradient boosted trees. Note: epoch for the tree-based model corresponds to the number of decision trees.}
\label{fig:algcomp_batch_size_20}
\end{figure*}

We also checked the results against a much smaller batch size in Fig.~\ref{fig:algcomp_batch_size_20}, in this case the batch size is 20 versus the 128 for Figs.~\ref{fig:bigfreakingplot}, \ref{fig:thebigone}, and \ref{fig:algcomp_ratio}, simply comparing QA (at the same number of Gibbs steps as used in the analyses in section \ref{sec:howtoboltzmann}) and discriminative training to logistic regression and gradient boosted trees. While performance improves and RBM structured models perform better than logistic regression, neither RBM model reaches the performance of gradient boosted trees, and no persistent advantage beyond a small number of training epochs is observed for QA over more efficient classical discriminative training. While resource and time limitations prohibited an exhaustive search or extensive comparison to hybrid algorithms, it is unlikely that the qualitative results discussed above would be effected.

\begin{figure}[htp]
\includegraphics[width=0.49\textwidth]{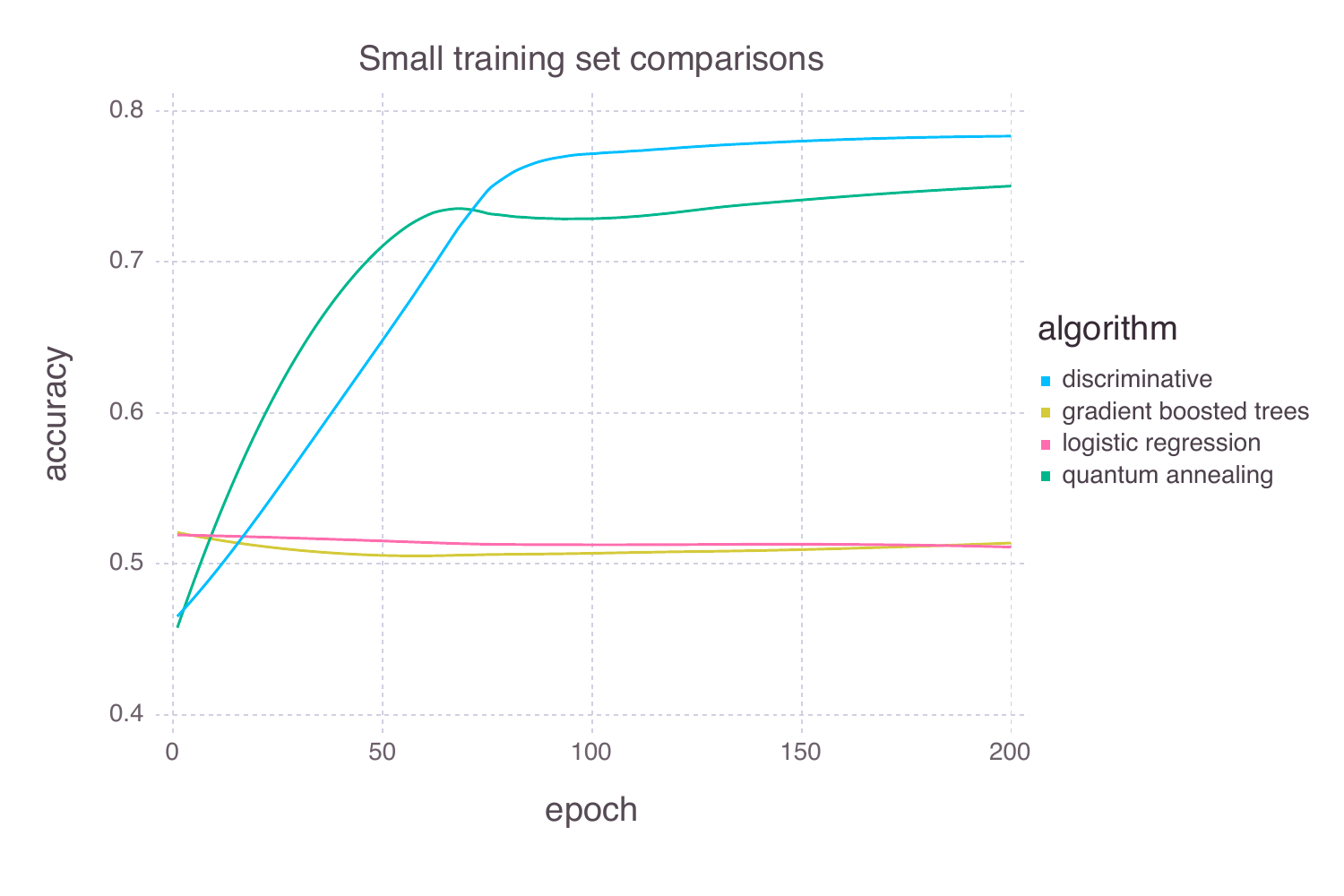}
\caption{Comparison of the different algorithms' accuracies on the test set when trained on only 250 training set examples.
We see that RBM beat the two algorithms we compare against, which both overfit the training set.
This is the case even for the gradient boosted classifier which is known as being robust to overfitting.
The two RBM training routines show a similar pattern to that of the larger training set. Batch size is 20, and QA is trained with 10 Gibbs steps of post-processing.}
\label{fig:smalltrainingset}
\end{figure}

Inspired by the observations in \cite{Mott2017}, and by the advantages shown very early in training by the QA approach, we also performed a study using a very small training set, with batch size of 20.
As shown in Fig.~\ref{fig:smalltrainingset}, the RBM approach actually shows a decisive advantage under this condition. 
When the training set is restricted to 250 events (reduced from 2500), we find strong overfitting when using logistic regression or gradient boosted trees, but good performance by the RBM.
We again see stronger performance by the QA (here with 10 Gibbs steps) early in the training process, but as the number of epochs of training increases, discriminatively trained models take over.

Curiously, this suggests that if there is a need to train a classifier based on very small datasets and if the number of epochs is limited (perhaps by time concerns in a situation where we were able to operate training for the QA resource with no network latency), then a QA-trained RBM shows promise as a superior algorithm.

\section{Conclusions and Future Work}
\label{sec:conclusions}

In this work, we explored a classification application of importance in cosmology, and studied the distribution of energy states coming from the D-Wave 2000Q.
We tested several post-processing methods that aimed to bring the output distributions closer to Boltzmann distributions, which are theoretically required for training RBMs, but we found little impact from post-processing on RBM performance.
As a consequence and for simplicity of interpretation, we subsequently minimized post-processing.
We presented the results of trained RBMs, and other ML algorithms, for galaxy morphology classification.
While we ultimately find RBMs implemented on D-Wave hardware perform well, we don't find compelling evidence of algorithmic performance advantage with this dataset for this problem over the most likely training scenarios.

We do not believe this result is an indictment of the performance of the quantum resources --- we found regions of phase space in the training where the quantum computer performed better than its classical counterpart.
In particular, for small datasets and for limited numbers of training repetitions, QA-based RBMs performed very well and outperformed the alternative classical algorithms studied here (logistic regression and gradient boosted trees), and outperformed classically trained RBMs.
However, outside of these rather special training scenarios, RBMs (regardless of the classical or quantum nature of the training algorithm) did not outperform the gradient boosted tree algorithm.

Perhaps more complex and higher dimensional data would be more challenging for algorithms like gradient boosted trees and regression, but here we find that they are able to handle the dataset well.
In the cases where significantly less or no compression is required due to significantly larger quantum resources, we may see a performance advantage for this algorithm.
This line of investigation will be interesting to revisit on future versions of quantum hardware, or perhaps on a digital annealer \cite{Inagaki2016,Aramon2019} with substantially larger RBMs.
For this data, due to the compression mechanisms involved, enlarging the RBM significantly did not lead to performance improvements, but with much larger numbers of qubits available, it would be possible to pursue different compression mechanisms or perhaps avoid compression altogether.
Another option to pursue with less compressed data would be increasing the network connectivity and adding additional layers.
There is evidence in classical machine learning \cite{alexnet} that multi-layer networks are able to construct hierarchical representations that often offer some advantage in data analysis tasks.

\begin{acknowledgments}
This manuscript has been authored by Fermi Research Alliance, LLC under Contract No. DE-AC02-07CH11359 with the U.S. Department of Energy, Office of Science, Office of High Energy Physics.
This research used resources of the Oak Ridge Leadership Computing Facility, which is a DOE Office of Science User Facility supported under Contract DE-AC05-00OR22725.
This project was funded in part by the DOE HEP QuantISED program.
S. Adachi and J. Job acknowledge Internal Research and Development funding from Lockheed Martin.
We thank D-Wave Systems for providing access to their DW2000Q systems.
We thank Travis Humble and Alex McCaskey for useful discussions about quantum machine learning, and Aristeidis Tsaris for useful comments and algorithms for the D-Wave.
We thank Maxwell Henderson, Carleton Coffrin and Vaibhaw Kumar for discussion on obtaining Boltzmann distributions from quantum annealers.
\end{acknowledgments}

\bibliographystyle{elsarticle-num-limitnames}
\bibliography{QuantumBib}

\end{document}